\documentclass[nolinenumbers]{aastex631}

\usepackage[T1]{fontenc}%
\usepackage{bm}
\usepackage{tabularx}
\usepackage{booktabs}
\usepackage{longtable,array,threeparttablex}
\usepackage{threeparttable}
\usepackage{ragged2e} 
\usepackage{booktabs,makecell, multirow, tabularx}
\usepackage{changepage}
\usepackage{amsmath} 

\usepackage{appendix}

\begin{document}

\title{A Photometric and Spectroscopic investigation of 11 TESS eclipsing contact binaries}

\author[0009-0008-3792-6444]{Yani Guo}
\affiliation{Shandong Key Laboratory of Space Environment and Exploration Technology, School of Space Science and Technology, Institute of Space Sciences, Shandong University, Weihai, Shandong, 264209, People's Republic of China}

\author[0000-0003-3590-335X]{Kai Li}
\affiliation{Shandong Key Laboratory of Space Environment and Exploration Technology, School of Space Science and Technology, Institute of Space Sciences, Shandong University, Weihai, Shandong, 264209, People's Republic of China}

\author[0000-0003-4207-1694]{Yanke Tang}
\affiliation{Shandong Key Laboratory of Space Environment and Exploration Technology, College of Physics and Electronic information, Dezhou University, 566 West University Road, Decheng District, Dezhou 253023, China}

\author[0009-0009-6364-0391]{Xiang Gao}
\affiliation{Shandong Key Laboratory of Space Environment and Exploration Technology, School of Space Science and Technology, Institute of Space Sciences, Shandong University, Weihai, Shandong, 264209, People's Republic of China}

\author{Qiqi Xia}
\affiliation{Shandong Key Laboratory of Space Environment and Exploration Technology, School of Space Science and Technology, Institute of Space Sciences, Shandong University, Weihai, Shandong, 264209, People's Republic of China}

\author[0009-0005-0485-418X]{Liheng Wang}
\affiliation{Shandong Key Laboratory of Space Environment and Exploration Technology, School of Space Science and Technology, Institute of Space Sciences, Shandong University, Weihai, Shandong, 264209, People's Republic of China}

\author{Meng Guo}
\affiliation{Shandong Key Laboratory of Space Environment and Exploration Technology, School of Space Science and Technology, Institute of Space Sciences, Shandong University, Weihai, Shandong, 264209, People's Republic of China}

\correspondingauthor{Kai Li}
\email{kaili@sdu.edu.cn}



\begin{abstract}
By cross-matching the eclipsing binary catalog provided by \cite{2022ApJS..258...16P} with LAMOST  medium resolution spectra, we obtained 11 targets. 
Combining light and radial velocity curves analysis, we have derived accurate physical parameters for these 11 targets.
The results indicate that there are 3 deep contact binaries, 3 moderate ones, and 5 shallow ones. Among them, 3 targets exhibit the O'Connell effect, which is attributed to the presence of star-spot on the component's surface. One target is a low-mass ratio deep contact binary and may be contact binary merging candidates. The evolutionary status of these 11 targets was studied using the mass-luminosity and mass-radius relation diagrams.
Based on the O-C (Observed minus Calculated) analysis of 10 targets, we found that the orbital periods of 5 contact binaries show a long-term decreasing trend, likely due to the combined effects of mass transfer between the two components and loss of angular momentum. Meanwhile, the orbital periods of the other 4 stars are continuously increasing, which is attributed to mass transfer. 
Besides, the O-C curves of 3 targets show clear periodic changes, which might result from the Applegate mechanism or the light travel time effect.

\end{abstract}



\section{Introduction} \label{sec:intro}

Contact binary systems are a special class of binary star systems, composed of two stars that are either filling or overflowing their Roche lobes and sharing a common envelope. These systems are relatively common in the Milky Way, with the majority having orbital periods shorter than 0.5 days. Contact binaries typically form from close binary systems with short orbital separations, where the two stars gradually move closer due to angular momentum loss caused by stellar winds and eventually become a contact binary \citep{1988ASIC..241..345G, 1994ASPC...56..228B, 2002ApJ...575..461E, 2003MNRAS.342.1260Q, 2006AcA....56..347S, 2017RAA....17...87Q}.
Current research generally suggests that contact binaries are likely to evolve into a single star through mass transfer and angular momentum loss \citep{1995ApJ...444L..41R, 2005AJ....130.1206Q, 2007MNRAS.377.1635A, 2010MNRAS.405.2485J, 2016RAA....16...68Z, 2017PASJ...69...79L, 2022AJ....164..202L, 2024A&A...692L...4L}. However, the specific details of evolutionary process, such as the rate of mass transfer, the mechanisms of angular momentum loss, and the response of the stellar internal structure, remain a subject of considerable debate. Further research is needed to reveal the exact evolutionary pathways of these systems.

The O'Connell effect \citep{1951PRCO....2...85O,2003ChJAA...3..142L,2009PASP..121.1366L,2013NewA...22...57L,2014ApJS..212....4Q,2014NewA...30...64L,2015NewA...41...17L, 2017PASJ...69...79L, 2016AJ....151...67Z} is a significant photometric phenomenon observed in contact binary systems, characterized by asymmetry in the light curve between the two eclipses, resulting in unequal maxima. This effect is typically associated with stellar magnetic activity, such as starspots or flares, which cause local changes in the stellar surface temperature and brightness, thereby affecting the shape of the light curve. The O’Connell effect has been detected in numerous contact binaries, serving as a vital clue for investigating stellar magnetic activity and internal structures of binary systems. Such research not only improves our understanding of the structural properties and evolutionary trajectories of contact binaries, but also clarifies how stellar magnetic activity influences these systems. 


With the rapid advancement of modern astronomy, survey projects have grown increasingly pivotal in astronomical research. Transiting Exoplanet Survey Satellite (TESS; \citealt{2015JATIS...1a4003R}) and Guoshoujing Telescope (the Large Sky Area Multi-Object Fiber Spectroscopic Telescope, LAMOST) \citep{2015RAA....15.1095L} are two highly representative projects. TESS, through its high-precision photometric observations, is capable of detecting minute changes in stellar brightness, providing a wealth of high-quality data for the study of light curves of stars. LAMOST, on the other hand, boasts strong spectroscopic observational capabilities, enabling the simultaneous acquisition of spectral information from a large number of celestial objects, offering robust support for the spectral analysis and determination of physical parameters of binary systems. When data from TESS and LAMOST are combined, photometric and spectroscopic information can complement each other. The high-precision photometric data from TESS can accurately depict the light curves of binary star systems, while the spectroscopic data from LAMOST can provide temperature, metallicity, and radial velocity, etc. This combination can not only derive precise physical parameters but also allows for a more precise study of their internal structure and evolutionary processes. This study presents the accurate physical parameters of 11 target systems through the joint analysis of their TESS photometric data and LAMOST spectroscopic data.

\begin{table}
\begin{adjustwidth}{-3cm}{0cm} 
\centering
\caption{Information of the 11 Targets.}
\label{tab:basic information}
\begin{threeparttable}
\begin{tabular}{ccccccccccc}
\hline
\hline
Targets & R.A. & Dec & Period(d) & $BJD_{0}$ & $T_{LAMOST}$ &  $T_{\text{TESS}}$ & $T_{\text{Gaia}}$(K) & $T_{\text{mean}}$ & Tmag & Sector\\
(TIC)& & & (d)&$(2457000+)$  & (K) & (K)& (K)&(K)  &\\
\midrule
16617827 & 20 31 08.79 & 38 47 00.5 & 0.290637 & 2420.18077 & ... & 5315 & ... & 5315 & 10.859      &  75  \\ 
20212631 & 15 08 11.38 & 39 58 12.9 & 0.297160 & 2717.48226  & 5588 & 5665 & 5789 & 5681 & 10.064   &  23  \\ 
116484192 & 15 33 24.85 & 45 34 07.4 & 0.276474 & 3395.73012  & 5518 & 5690 & 5644 & 5617 & 11.875  &  77  \\ 
122712115 & 19 30 22.96 & 40 55 01.2 & 0.731494 & 1697.29700  & 7553 & 7554 & 7625 & 7577 & 11.493  &  40  \\ 
142587827 & 11 12 18.24 & 73 06 55.0 & 0.439398 & 2945.49309  & ... & 6696 & ... & 6696 & 8.417     &  75  \\ 
155968973 & 15 45 17.58 & 43 45 45.3 & 0.341881 & 3395.78564  & 5429 & 5383 & 5179 & 5330 & 11.989  &  51  \\ 
198410119 & 17 19 42.62 & 53 59 27.6 & 0.293898 & 1955.92482  & ... & 5613 & 5820 & 5717 & 11.055   &  60  \\ 
219109908 & 18 09 08.42 & 69 45 15.5 & 0.424027 & 2716.53075  & ... & 5937 & 5736 & 5836 & 10.42    &  73  \\ 
286169068 & 15 10 30.61 & 53 43 40.5 & 0.414619 & 1919.60201  & ... & 7204 & 7057 & 7131 & 12.124   &  77  \\ 
349294422 & 03 32 15.38 & 30 01 19.2 & 0.310388 & 2448.06237  & 5683 & 5669 & 5503 & 5618 & 10.683  &  71  \\ 
367683204 & 12 12 06.06 & 22 31 58.7 & 0.220685 & 1900.09600  & 4168 & 4148 & ... & 4158 & 10.431   &  48  \\ 
\hline
\end{tabular}
\end{threeparttable}
\end{adjustwidth}
\end{table}

\begin{table}
\begin{adjustwidth}{-3cm}{0cm} 
\centering
\caption{The Radial Velocity Values of the 11 Targets.}
\label{tab:RV}
\begin{threeparttable}
\begin{tabular}{ccccccccccc}
\hline
\hline
TIC 16617827 & BJD & Phase & RV-1           & Error           & RV-2           & Error \\
             &     &       &  (km $s^{-1}$) & (km $s^{-1}$)   & (km $s^{-1}$)  &(km $s^{-1}$)\\
\hline
&  2459868.93986 	&  0.056 &	-212.49 &	1.07  &	  107.87 	&  1.10 \\
&  2459870.97807 	&  0.069 &	-176.35 &	0.88  &	  77.24 	&  0.95 \\
&  2459122.00797 	&  0.074 &	-128.87 &	4.20  &	  42.09 	&  3.54 \\
&  2459151.94407 	&  0.076 &	-130.99 &	2.34  &	  37.85 	&  2.16 \\
&  2459870.99328 	&  0.122 &	-210.88 &	0.93  &	  96.94 	&  0.96 \\
&  2459122.02427 	&  0.130 &	-161.95 &	1.15  &	  61.92 	&  1.23 \\
&  2459151.96031 	&  0.132 &	-160.14 &	1.02  &	  65.74 	&  1.06 \\
&  2459871.00847 	&  0.174 &	-222.14 &	0.96  &	  110.71 	&  1.02 \\
&  2458409.98056 	&  0.189 &	-146.76 &	1.30  &	  57.44 	&  1.40 \\
&  2459122.04052 	&  0.186 &	-194.96 &	0.88  &	  85.44 	&  0.93 \\
&  2459151.97654 	&  0.188 &	-203.25 &	0.94  &	  81.26 	&  1.06 \\
&  2459868.97941 	&  0.192 &	-226.54 &	0.91  &	  111.96 	&  0.97 \\
&  2459871.02368 	&  0.226 &	-225.52 &	0.90  &	  108.89 	&  1.01 \\
&  2458409.99680 	&  0.244 &	-188.05 &	1.00  &	  74.59 	&  1.05 \\
&  2459868.99465 	&  0.245 &	-220.50 &	0.95  &	  107.13 	&  1.01 \\
&  2459124.97336 	&  0.277 &	-228.36 &	1.19  &	  103.72 	&  1.20 \\
&  2458410.01307 	&  0.300 &	-211.96 &	0.87  &	  91.80 	&  0.99 \\
&  2459869.00987 	&  0.297 &	-193.79 &	0.95  &	  88.86 	&  1.13 \\
&  2459124.98959 	&  0.333 &	-215.66 &	0.95  &	  96.74 	&  1.00 \\
&  2459869.02509 	&  0.350 &	-150.98 &	1.09  &	  78.95 	&  1.27 \\
&  2458410.02928 	&  0.356 &	-210.48 &	1.81  &	  111.48 	&  1.42 \\
&  2459125.00585 	&  0.389 &	-193.53 &	0.93  &	  77.29 	&  1.09 \\
&  2459368.31684 	&  0.554 &	 66.32 	& 5.41  &	  -93.23 	&  6.45 \\
&  2459127.99211 	&  0.664 &	 154.39 &	0.99  &	  -111.83 &	 1.02 \\
&  2458410.99837 	&  0.691 &	 81.30 	& 1.72  &	  -118.27 &	 1.74 \\
&  2459128.00838 	&  0.720 &	 198.76 &	1.08  &	  -117.93 &	 1.04 \\
&  2458411.01462 	&  0.746 &	 148.53 &	1.11  &	  -116.29 &	 1.15 \\
&  2459128.02463 	&  0.776 &	 213.34 &	0.97  &	  -124.63 &	 1.10 \\
&  2458415.98111 	&  0.835 &	 207.47 &	1.13  &	  -122.80 &	 1.26 \\
&  2458415.99734 	&  0.891 &	 216.68 &	0.95  &	  -119.34 &	 1.01 \\
&  2458416.01357 	&  0.946 &	 193.76 &	1.03  &	  -103.56 &	 1.14 \\
&  2459870.94574 	&  0.958 &   ...    &  ...  &     -7.62 &	 0.68 \\
\hline
\end{tabular}
\begin{tablenotes}  
\item \hspace{2.2cm} \textbf{Note--} This table is available in its entirety in machine-readable form. 
\end{tablenotes} 
\end{threeparttable}
\end{adjustwidth}
\end{table}

\section{Observations} \label{sec:style}

\subsection{Target Selection and Data Processing}

Since its launch in 2018, TESS has provided a large amount of high-precision light curve data for astronomical research. The design of TESS includes four wide-angle cameras, each with a field of view of $24^{\circ}$ × $24^{\circ}$. TESS divides the sky into 26 observation sectors, with each sector being observed for about 27 days, thus providing a large amount of data.

LAMOST is an active optics reflecting Schmidt telescope with a field of view of 5 degrees. The optical system of the telescope comprises a reflective Schmidt correcting plate $M_{A}$ with an aperture of 5.7 meters × 4.4 meters, a spherical primary mirror $M_{B}$ with an aperture of 6.7 meters × 6 meters, and a focal plane with a diameter of 1.75 meters. On the focal plane within the 5-degree field of view, LAMOST is equipped with 4,000 fibers, enabling it to obtain spectra of up to 4,000 celestial objects simultaneously in a single observation, making it the telescope with one of the highest spectral acquisition rate in the world. In the Medium-Resolution Spectroscopic Survey (MRS) mode, the blue arm of LAMOST covers a wavelength range of 4950 \AA to 5350 \AA, while the red arm covers 6300  \AA to 6800 \AA, with a spectral resolution of approximately $R \approx 7500$.

\cite{2022ApJS..258...16P} identified 4584 eclipsing binary stars from the 2-minute cadence data in sectors 1-26 during the first two years of the TESS mission and provided a catalog website\footnote[1]{https://tessebs.villanova.edu/}. 
We selected targets from this catalog and cross-matched them with LAMOST MRS. 
We select targets with a signal-to-noise ratio greater than 10 (SNR > 10) and more than six observations from the MRS data of LAMOST DR12. Finally, a total of 11 targets were identified. The detailed information of these 11 targets is listed in Table \ref{tab:basic information}.

\begin{figure*}
    \centering
    \begin{minipage}[t]{0.24\textwidth}
        \centering
        \includegraphics[width=\linewidth]{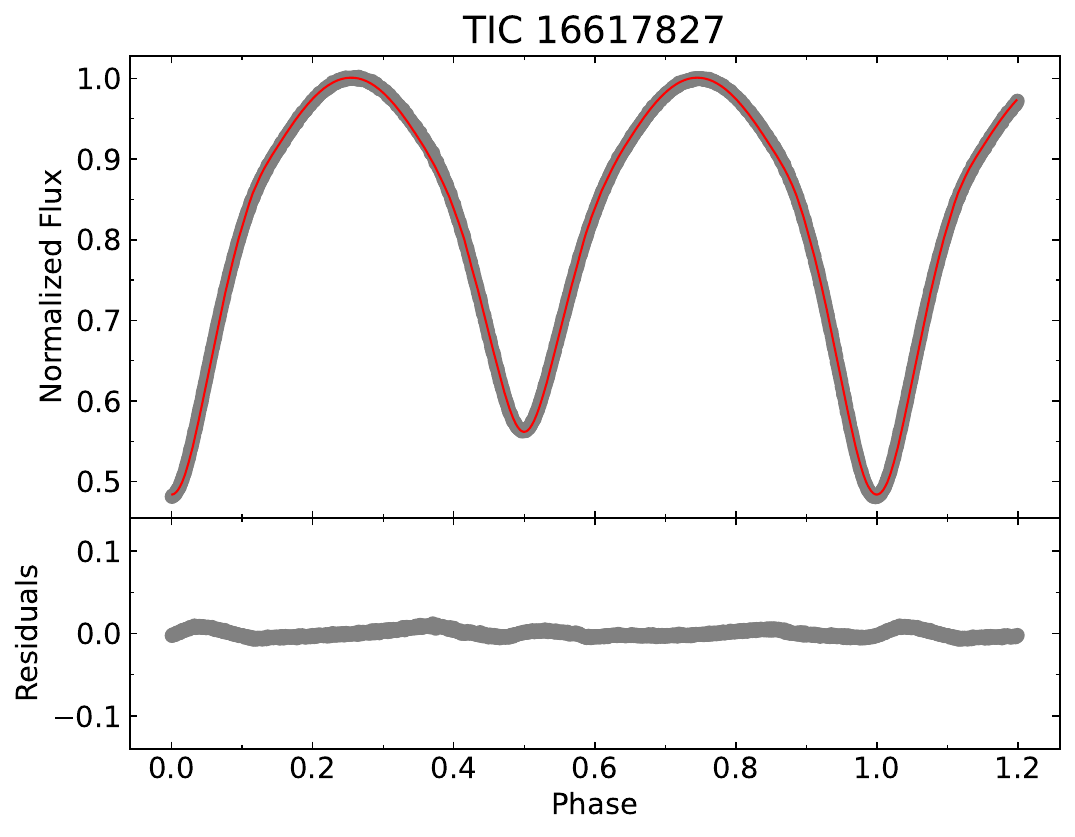}
    \end{minipage}
    \hfill
    \begin{minipage}[t]{0.24\textwidth}
        \centering
        \vspace{-3.4cm} 
        \includegraphics[width=\linewidth]{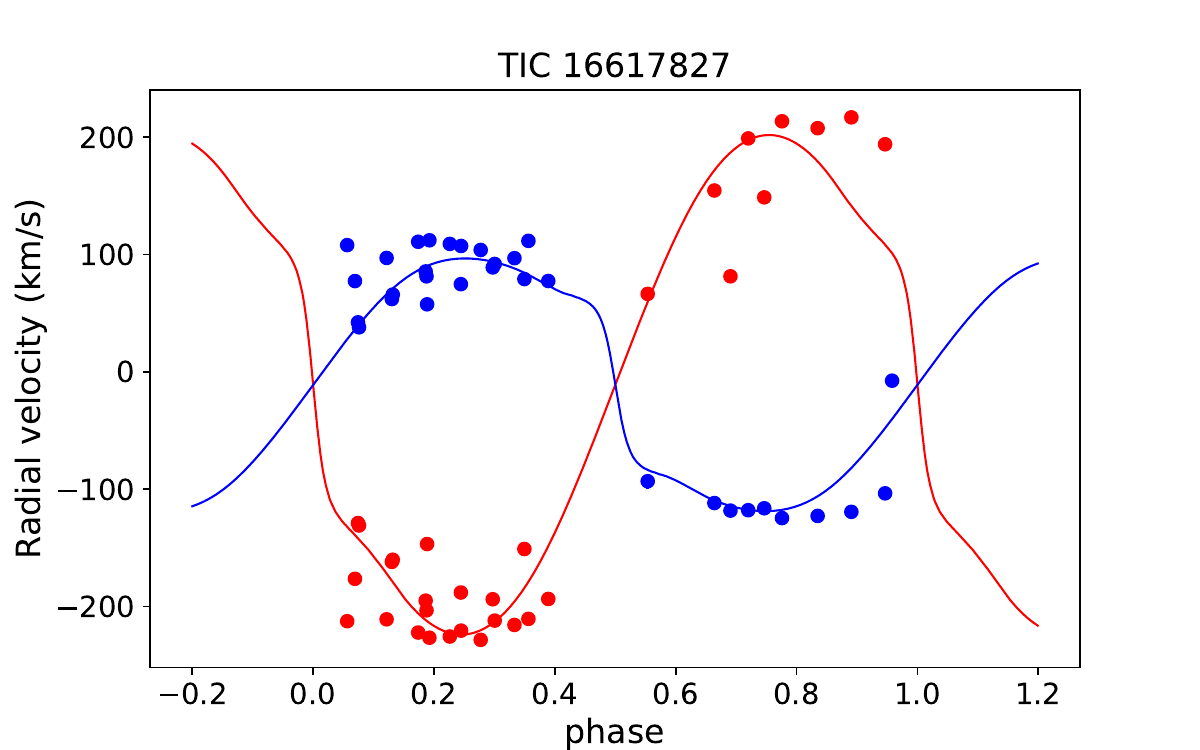}
    \end{minipage}
    \hfill
    \begin{minipage}[t]{0.24\textwidth}
        \centering
        \includegraphics[width=\linewidth]{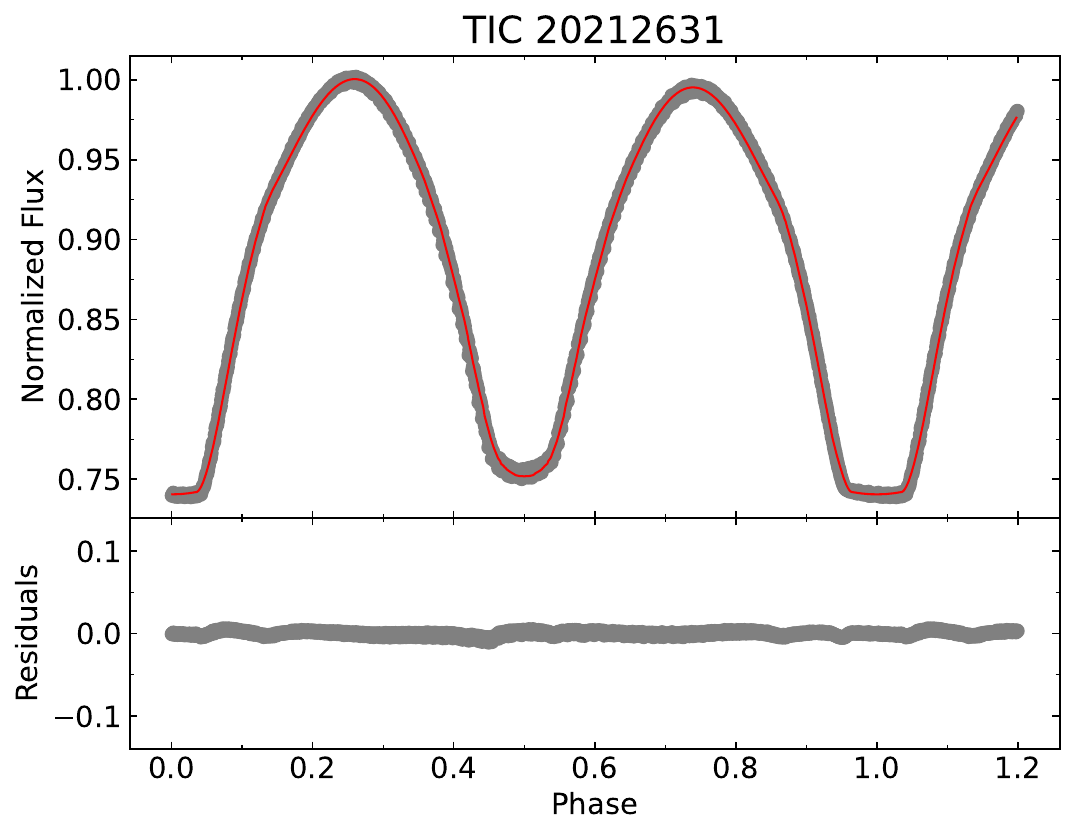}
    \end{minipage}
    \hfill
    \begin{minipage}[t]{0.24\textwidth}
        \centering
        \vspace{-3.4cm} 
        \includegraphics[width=\linewidth]{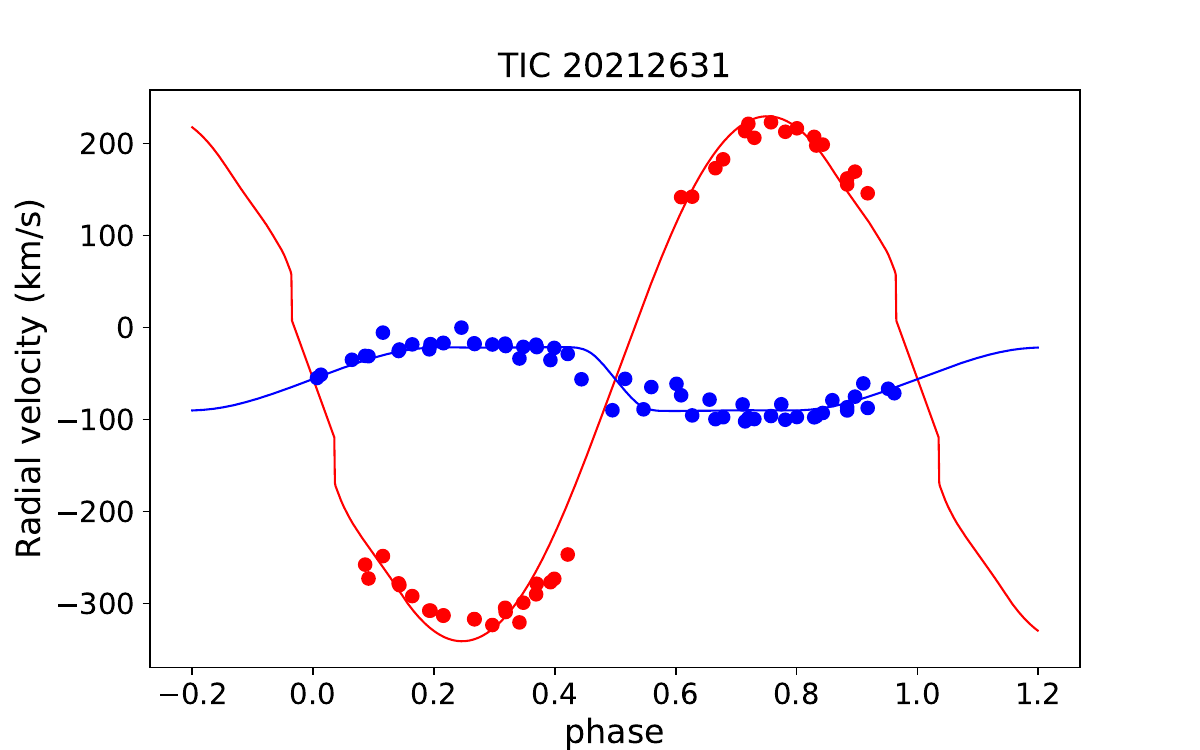}
    \end{minipage}
    
    \begin{minipage}[t]{0.24\textwidth}
        \centering
        \includegraphics[width=\linewidth]{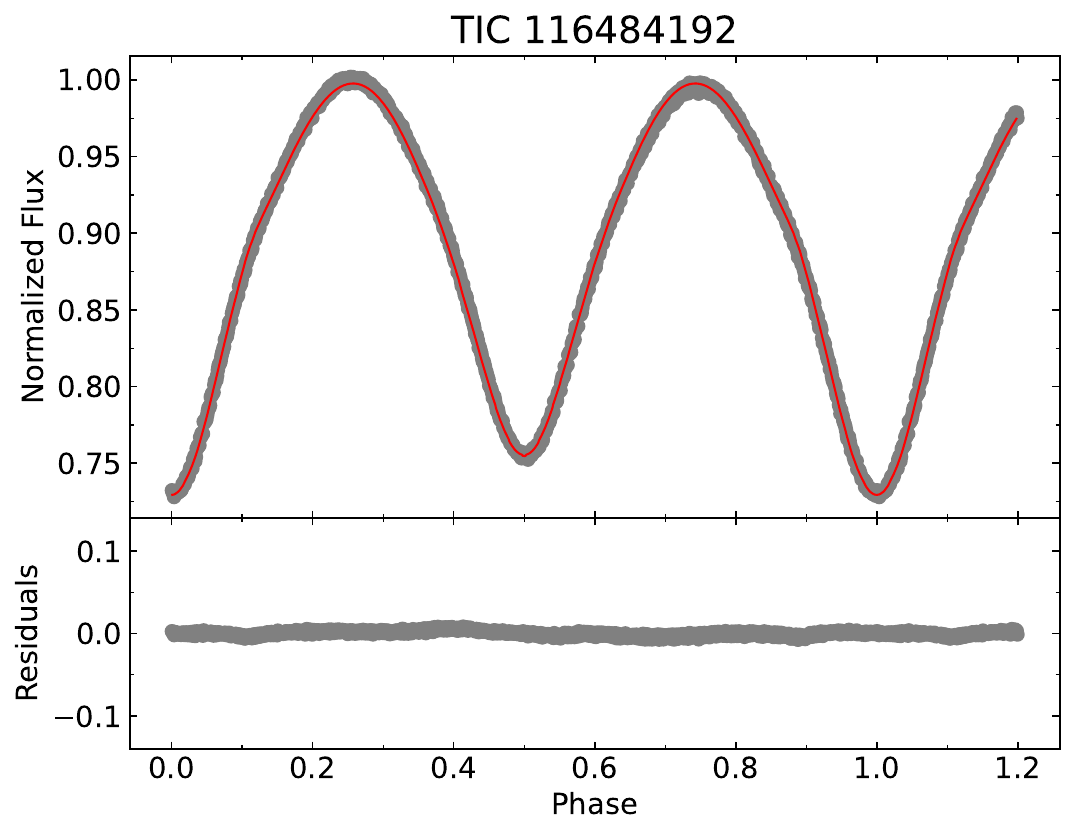}
    \end{minipage}
    \hfill
    \begin{minipage}[t]{0.24\textwidth}
        \centering
        \vspace{-3.4cm} 
        \includegraphics[width=\linewidth]{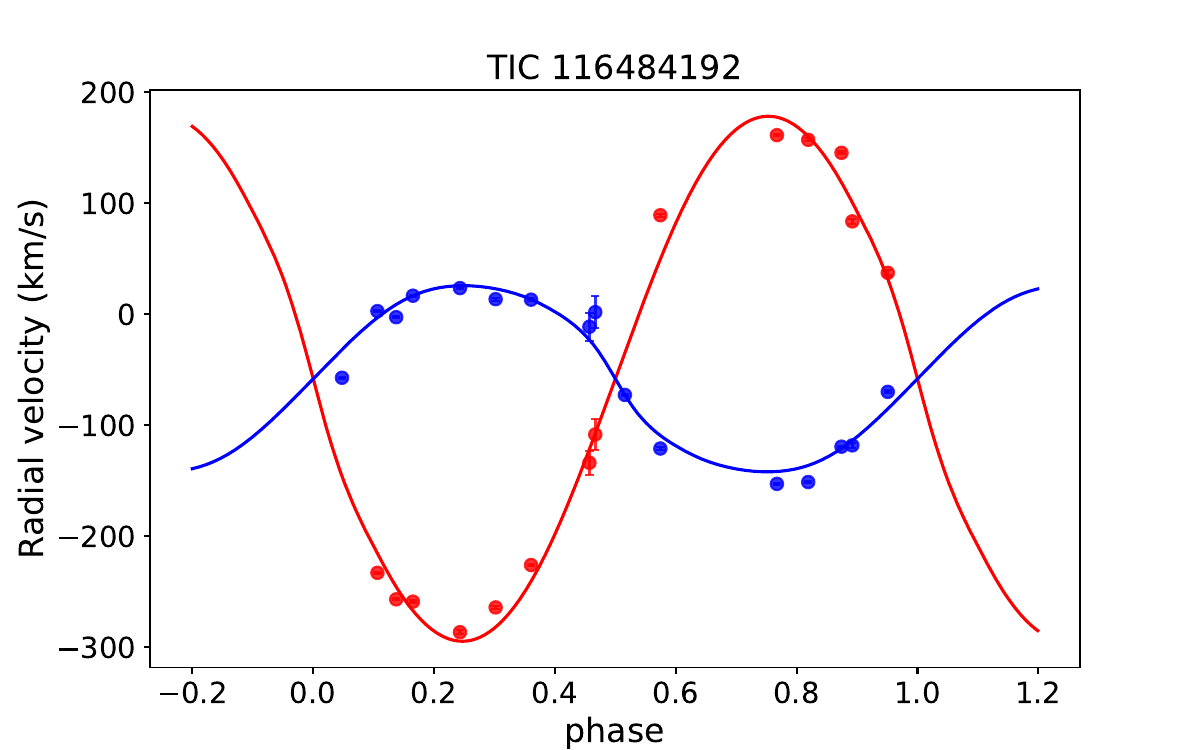}
    \end{minipage}
    \hfill
    \begin{minipage}[t]{0.24\textwidth}
        \centering
        \includegraphics[width=\linewidth]{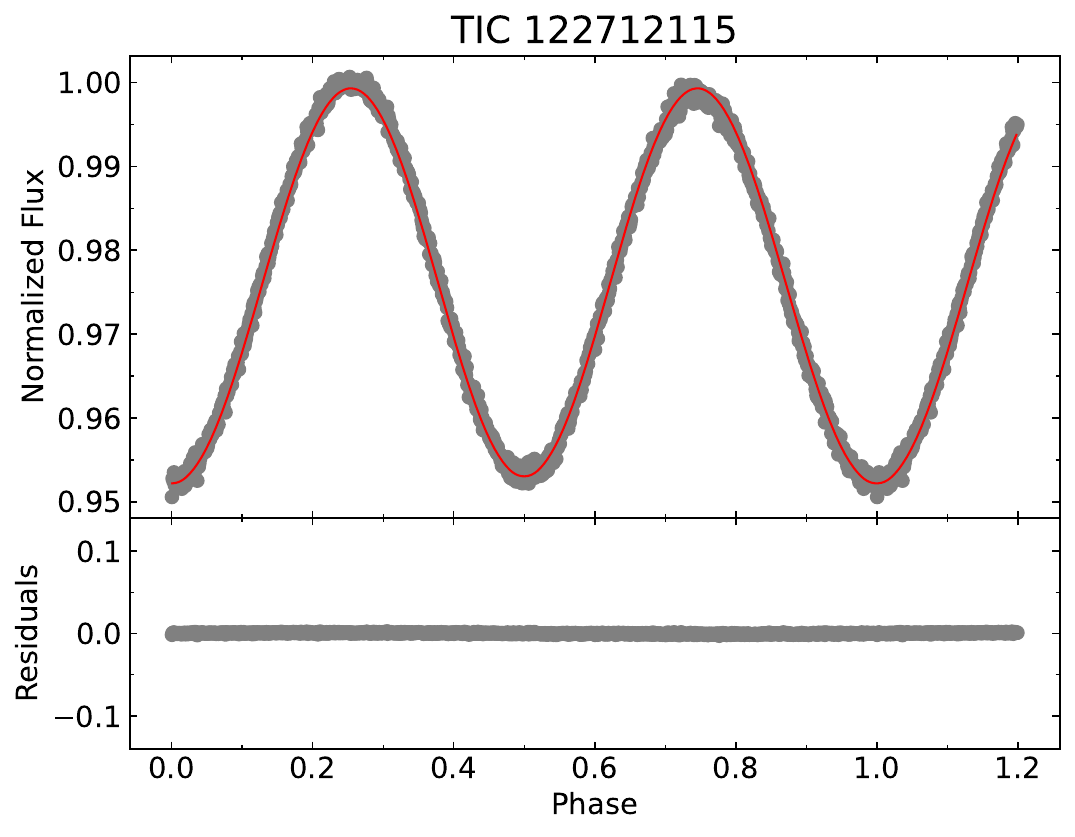}
    \end{minipage}
    \hfill
    \begin{minipage}[t]{0.24\textwidth}
        \centering
        \vspace{-3.4cm} 
        \includegraphics[width=\linewidth]{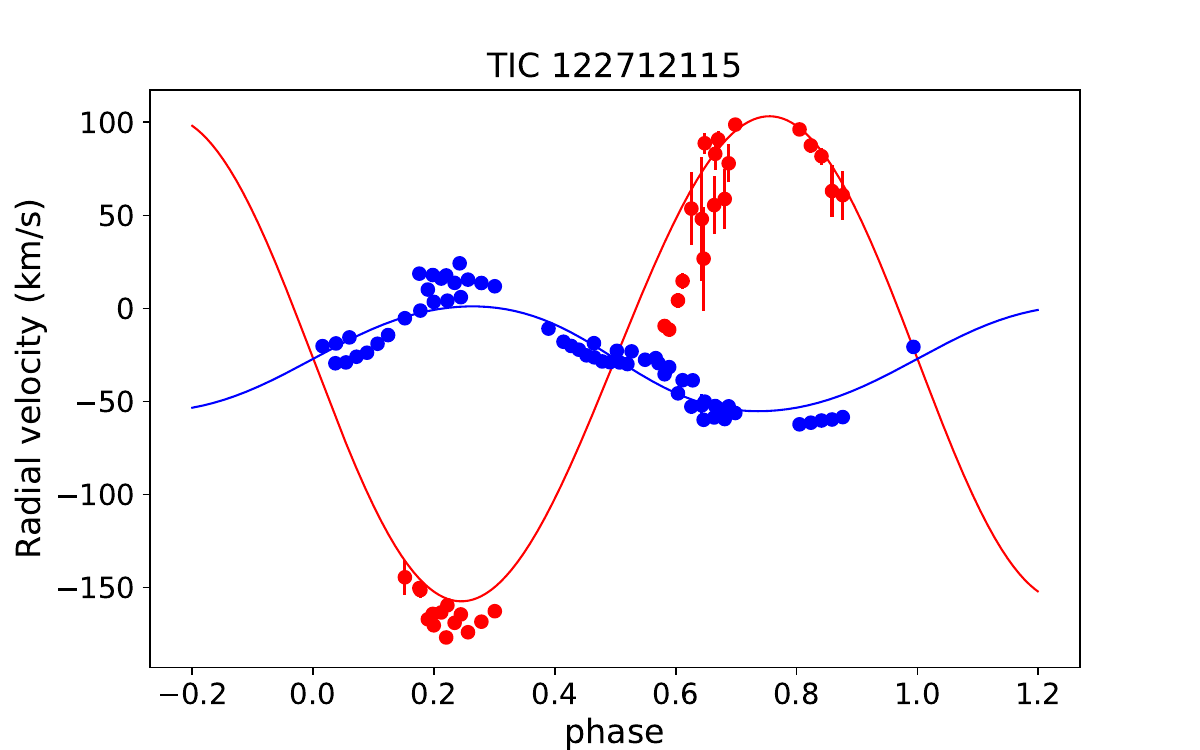}
    \end{minipage}
    
    \begin{minipage}[t]{0.24\textwidth}
        \centering
        \includegraphics[width=\linewidth]{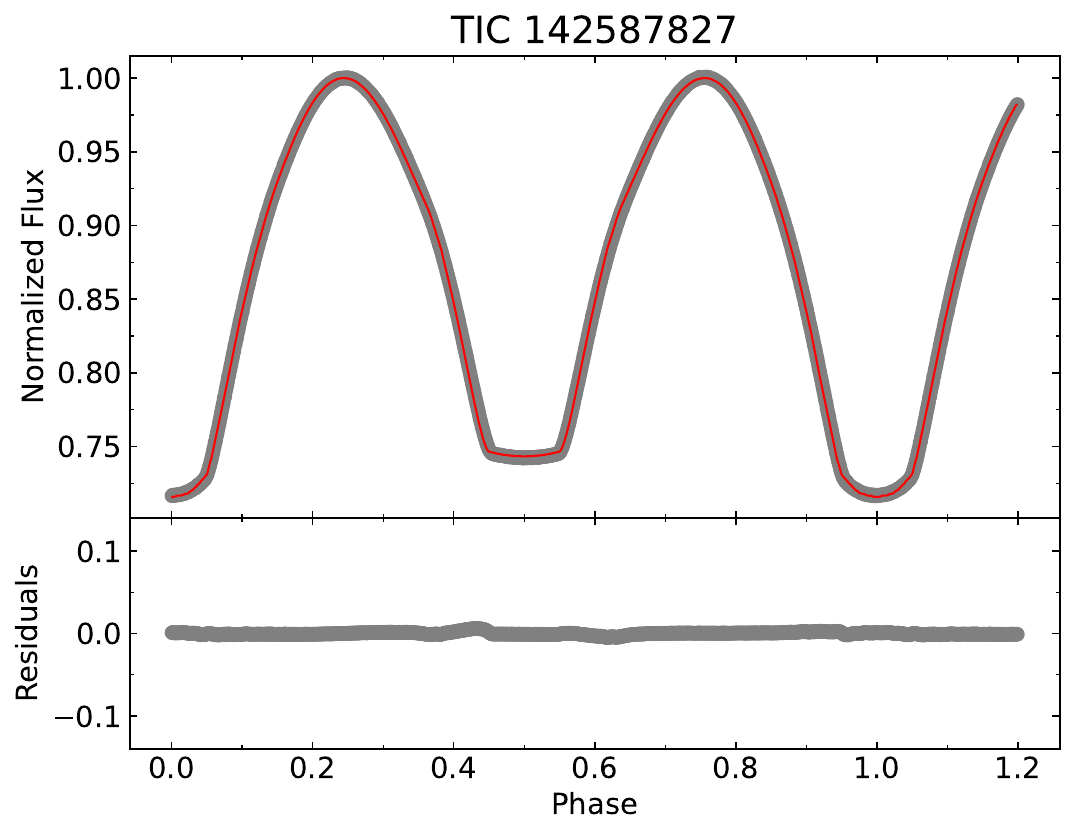}
    \end{minipage}
    \hfill
    \begin{minipage}[t]{0.24\textwidth}
        \centering
        \vspace{-3.4cm} 
        \includegraphics[width=\linewidth]{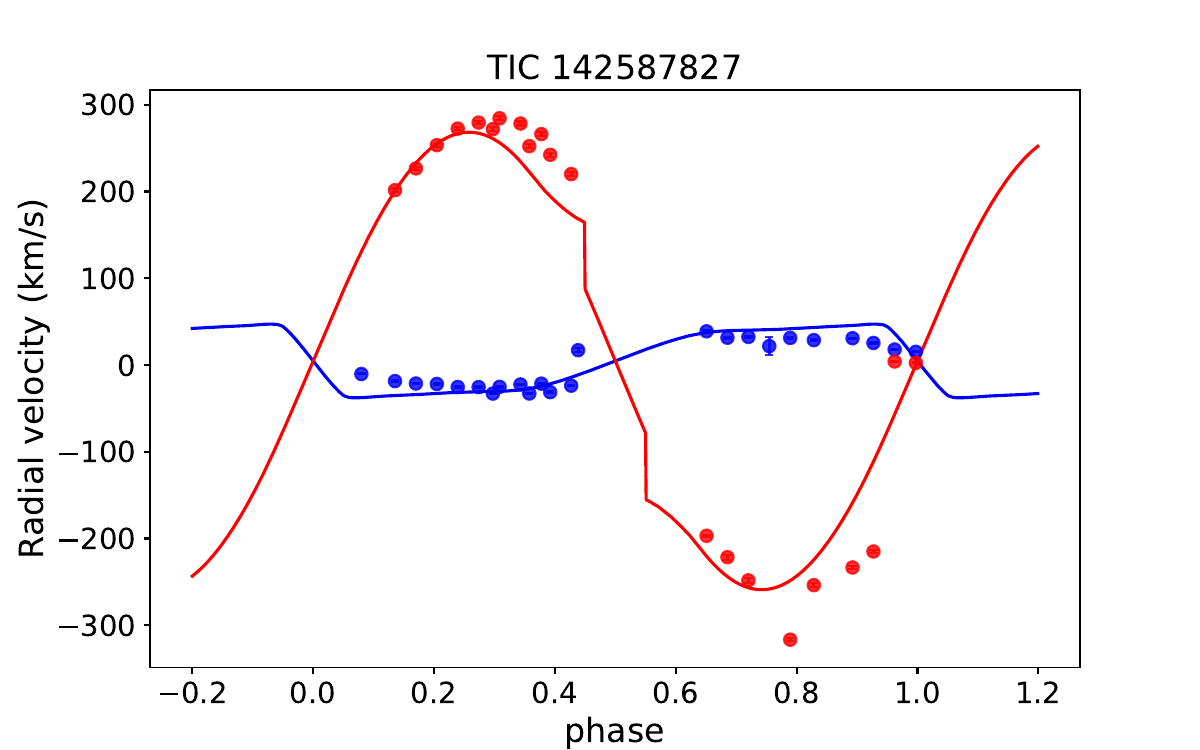}
    \end{minipage}
    \hfill
    \begin{minipage}[t]{0.24\textwidth}
        \centering
        \includegraphics[width=\linewidth]{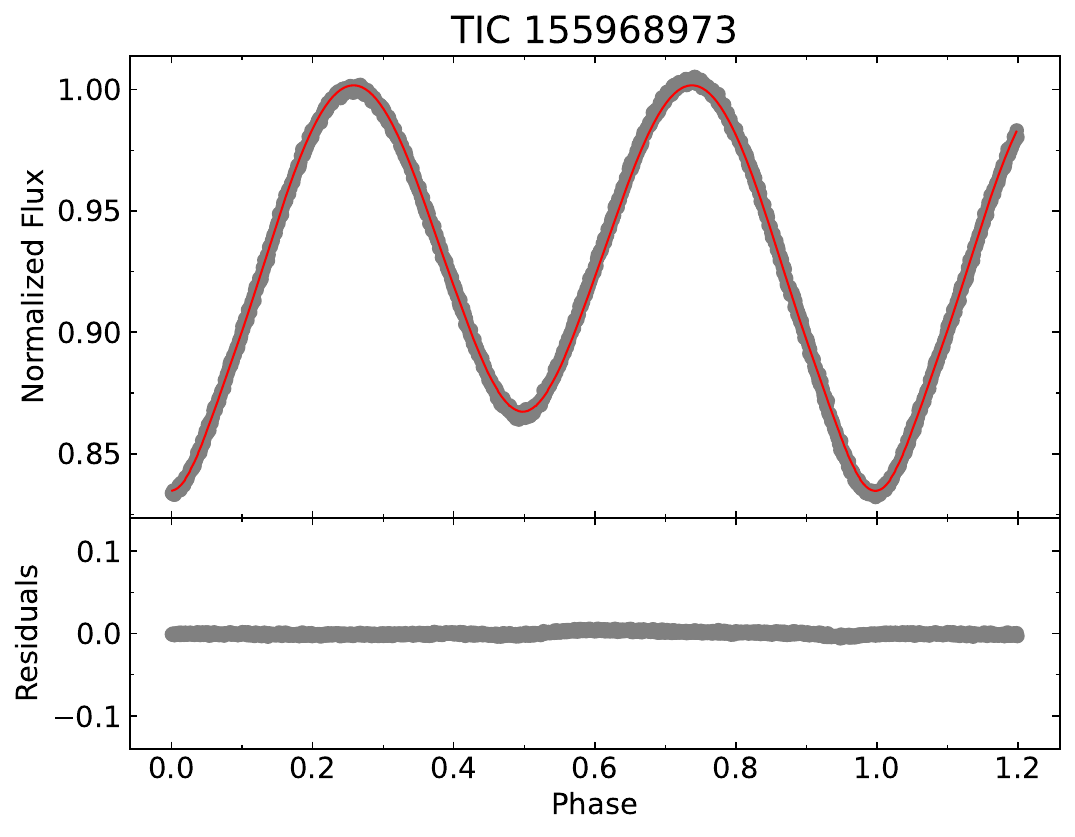}
    \end{minipage}
    \hfill
    \begin{minipage}[t]{0.24\textwidth}
        \centering
        \vspace{-3.4cm} 
        \includegraphics[width=\linewidth]{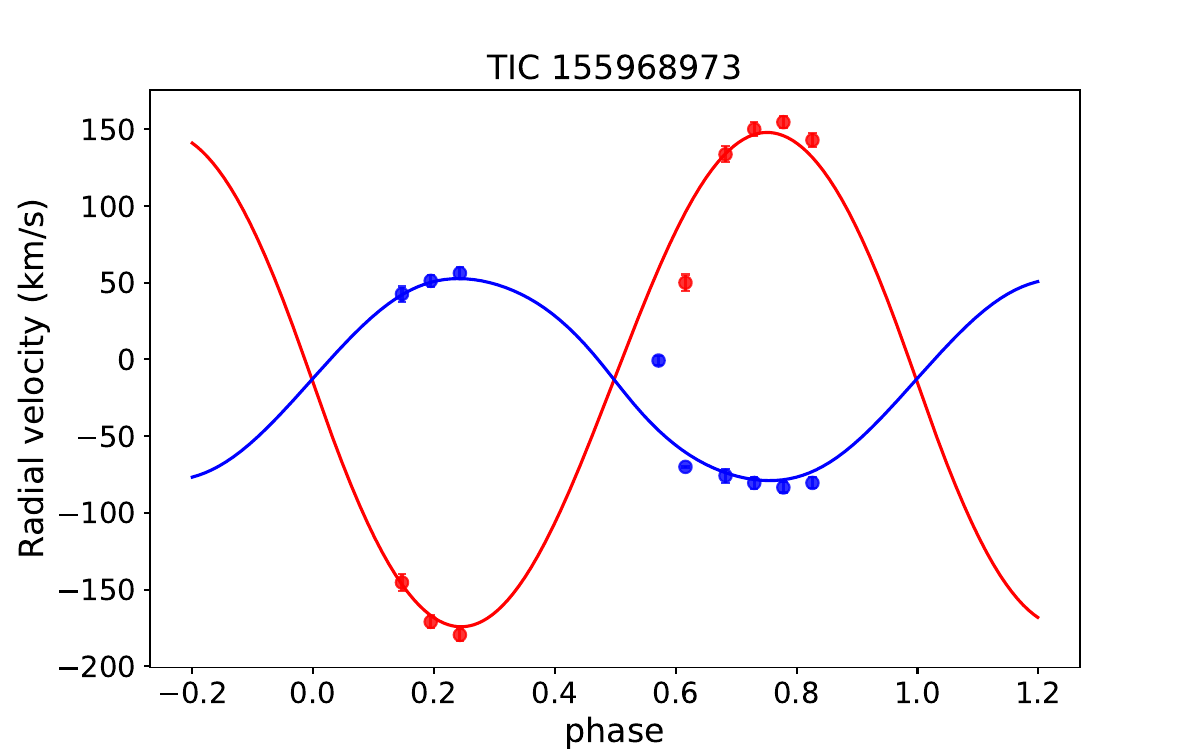}
    \end{minipage}

    \begin{minipage}[t]{0.24\textwidth}
        \centering
        \includegraphics[width=\linewidth]{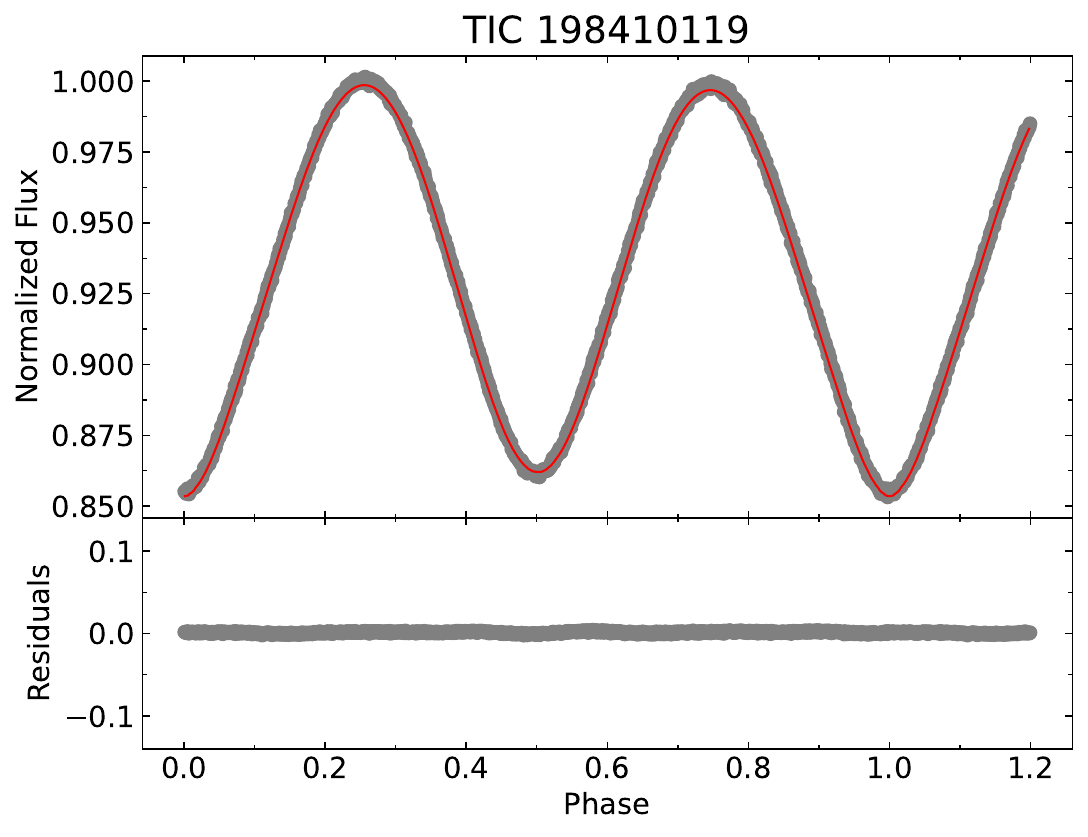}
    \end{minipage}
    \hfill
    \begin{minipage}[t]{0.24\textwidth}
        \centering
        \vspace{-3.4cm} 
        \includegraphics[width=\linewidth]{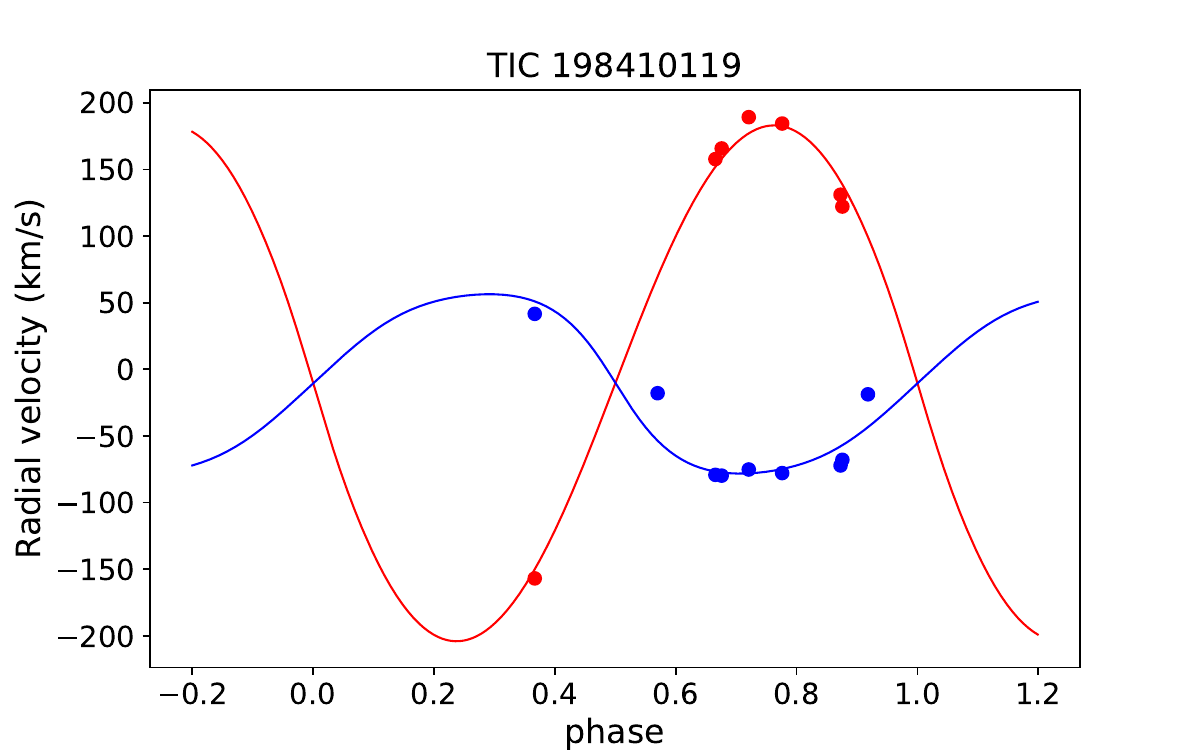}
    \end{minipage}
    \hfill
    \begin{minipage}[t]{0.24\textwidth}
        \centering
        \includegraphics[width=\linewidth]{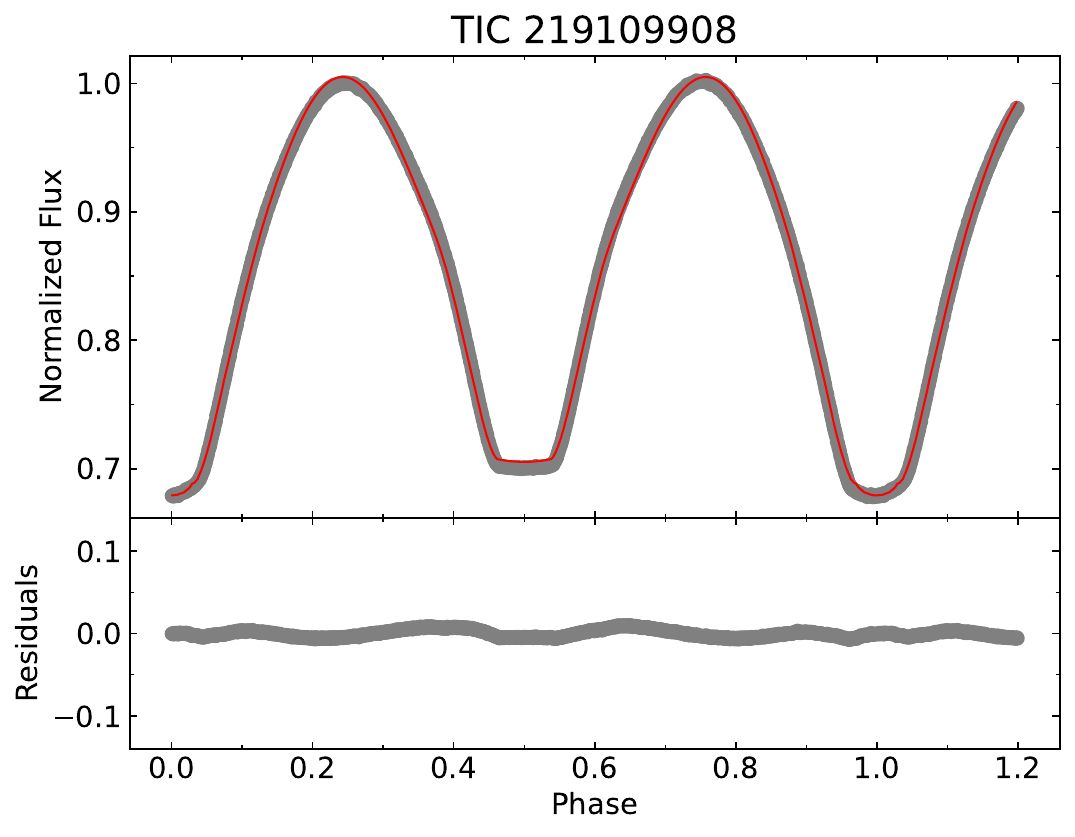}
    \end{minipage}
    \hfill
    \begin{minipage}[t]{0.24\textwidth}
        \centering
        \vspace{-3.4cm} 
        \includegraphics[width=\linewidth]{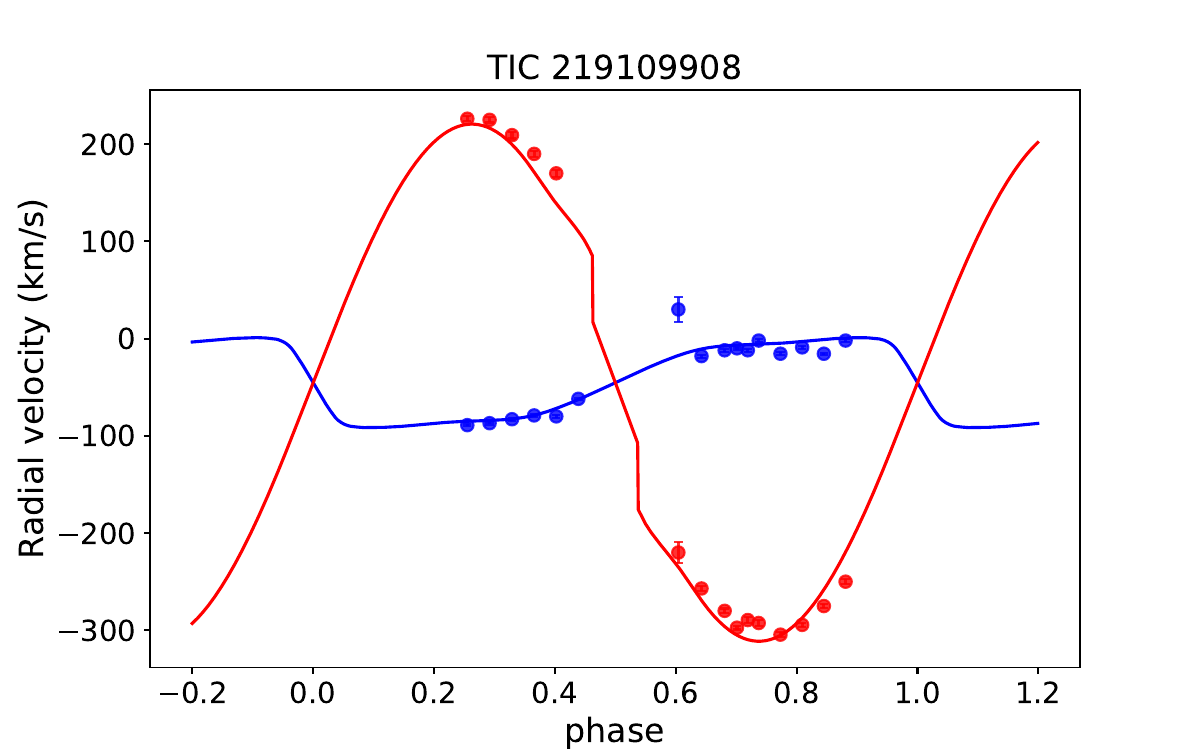}
    \end{minipage}

    \begin{minipage}[t]{0.24\textwidth}
        \centering
        \includegraphics[width=\linewidth]{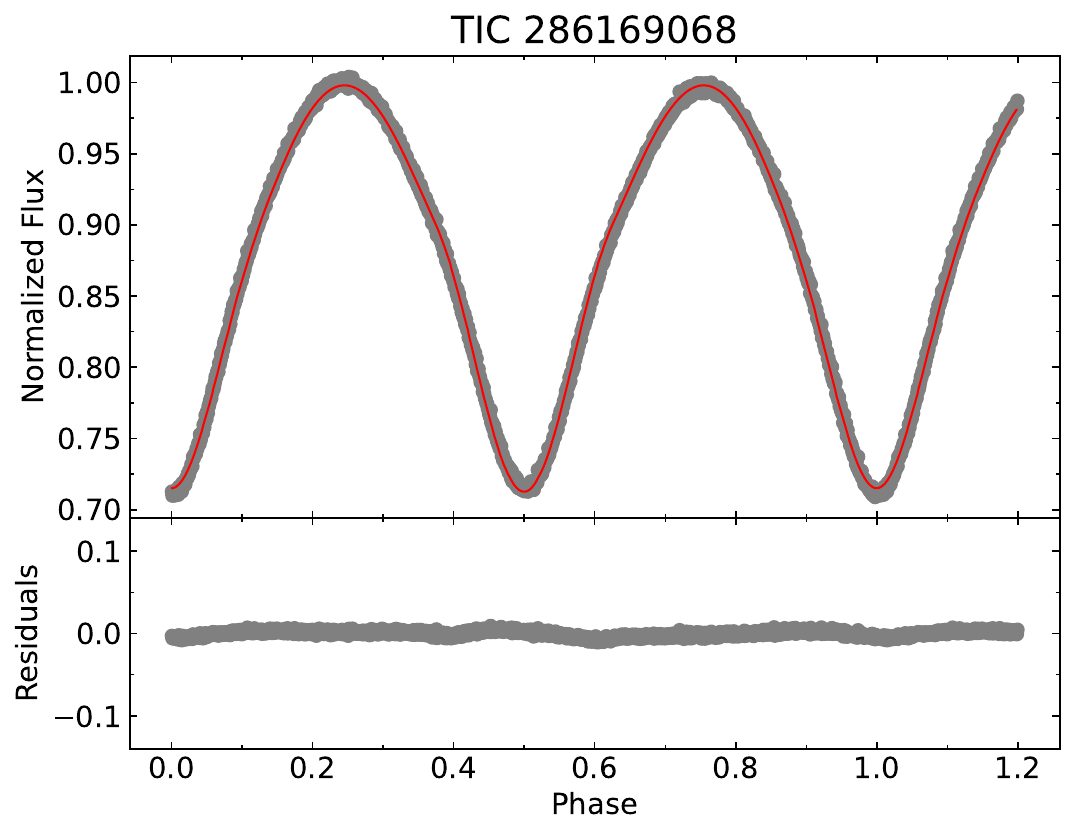}
    \end{minipage}
    \hfill
    \begin{minipage}[t]{0.24\textwidth}
        \centering
        \vspace{-3.4cm} 
        \includegraphics[width=\linewidth]{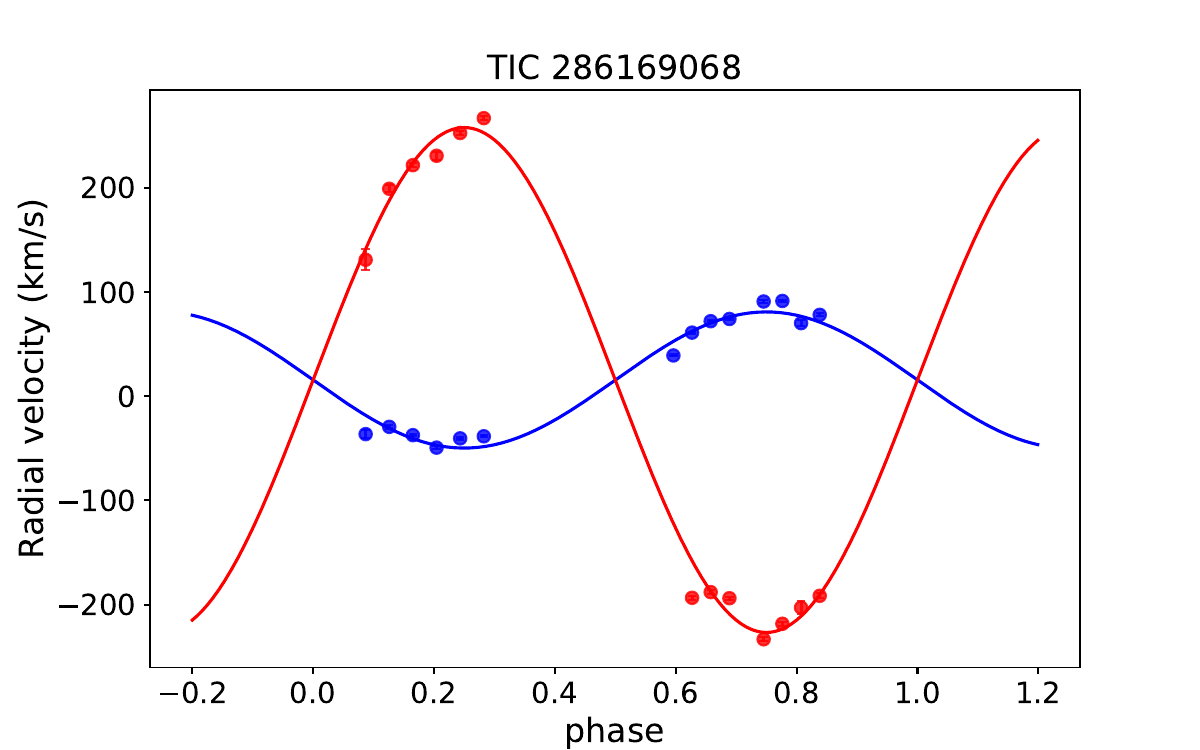}
    \end{minipage}
    \hfill
    \begin{minipage}[t]{0.24\textwidth}
        \centering
        \includegraphics[width=\linewidth]{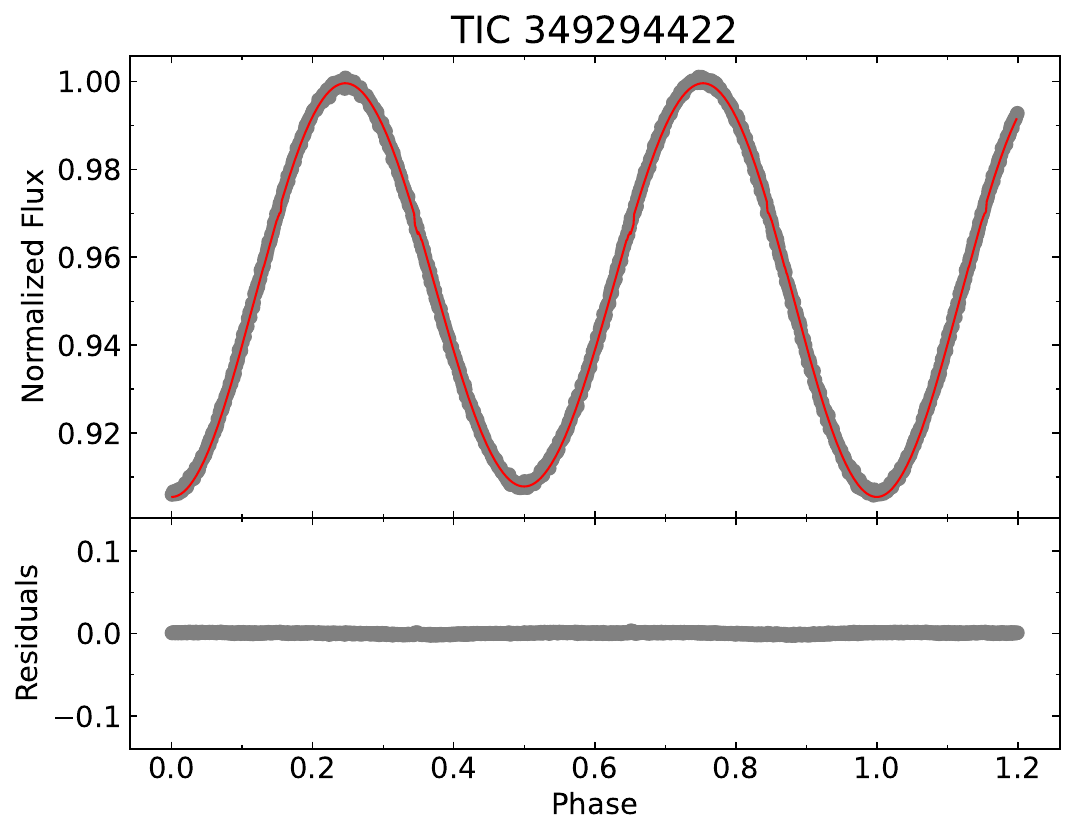}
    \end{minipage}
    \hfill
    \begin{minipage}[t]{0.24\textwidth}
        \centering
        \vspace{-3.4cm} 
        \includegraphics[width=\linewidth]{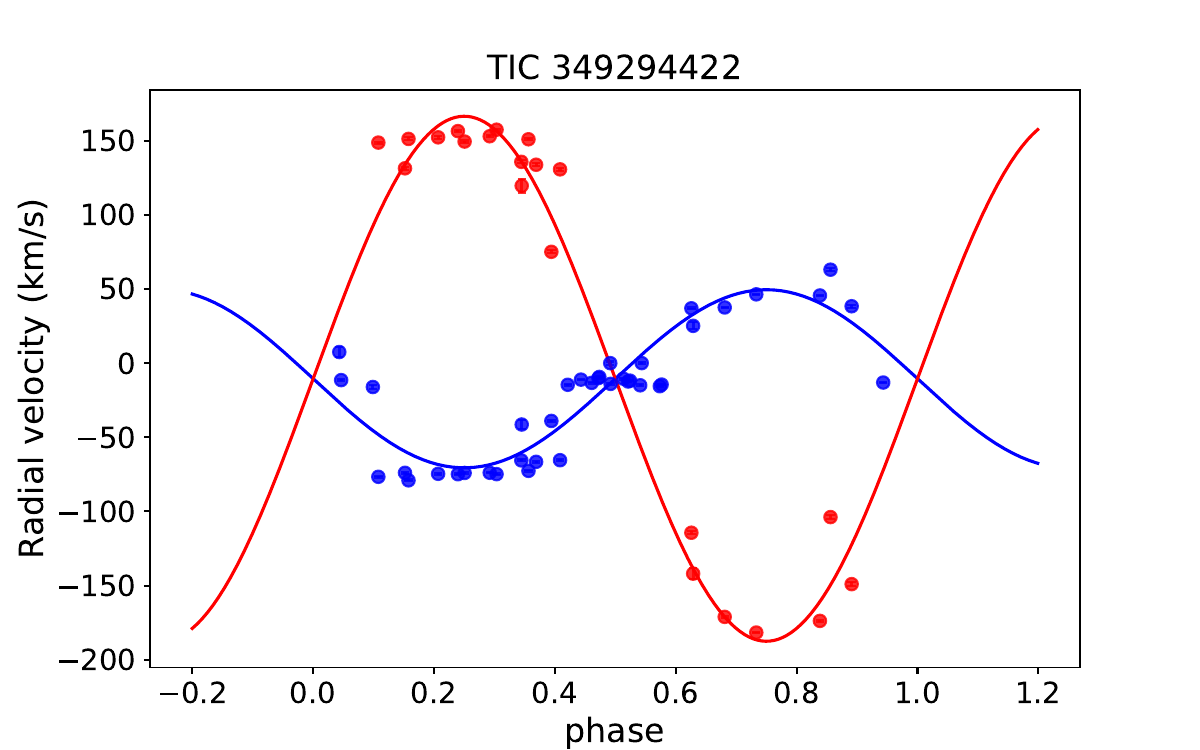}
    \end{minipage}
    
    \begin{minipage}[t]{0.24\textwidth}
        \centering
        \includegraphics[width=\linewidth]{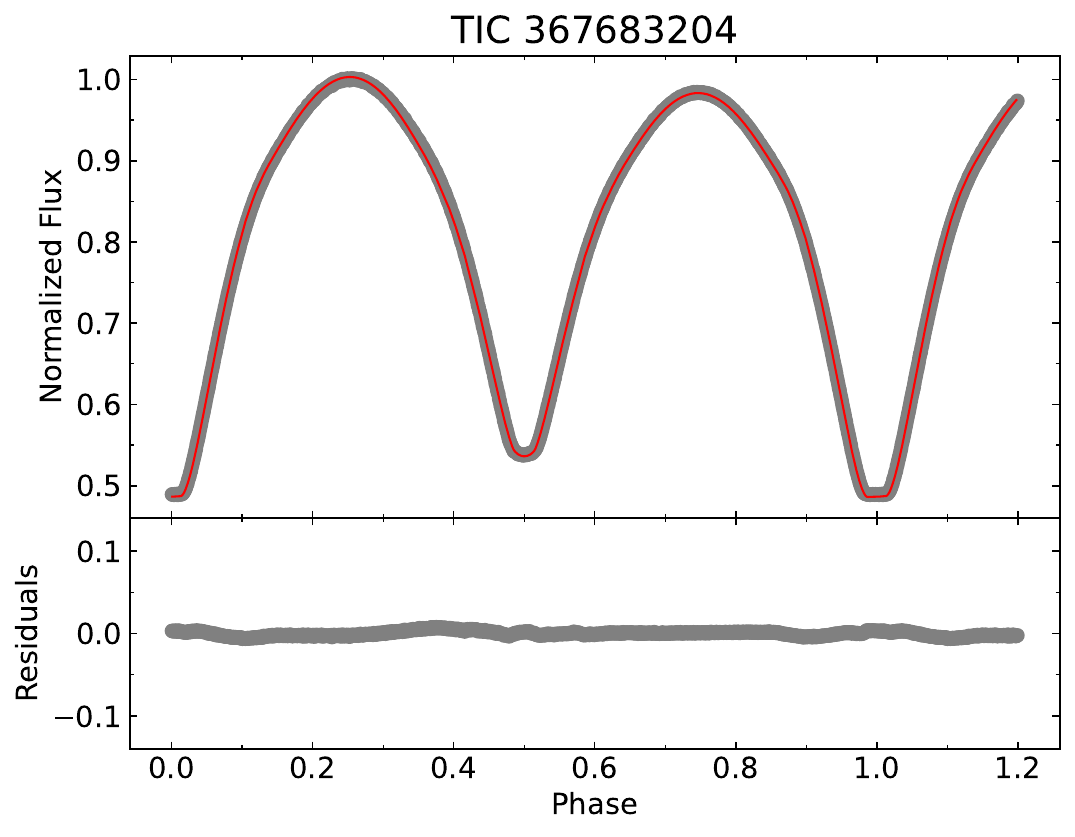}
    \end{minipage}
    \hfill
    \begin{minipage}[t]{0.24\textwidth}
        \centering
        \vspace{-3.4cm} 
        \hspace{-18.3cm} 
        \includegraphics[width=\linewidth]{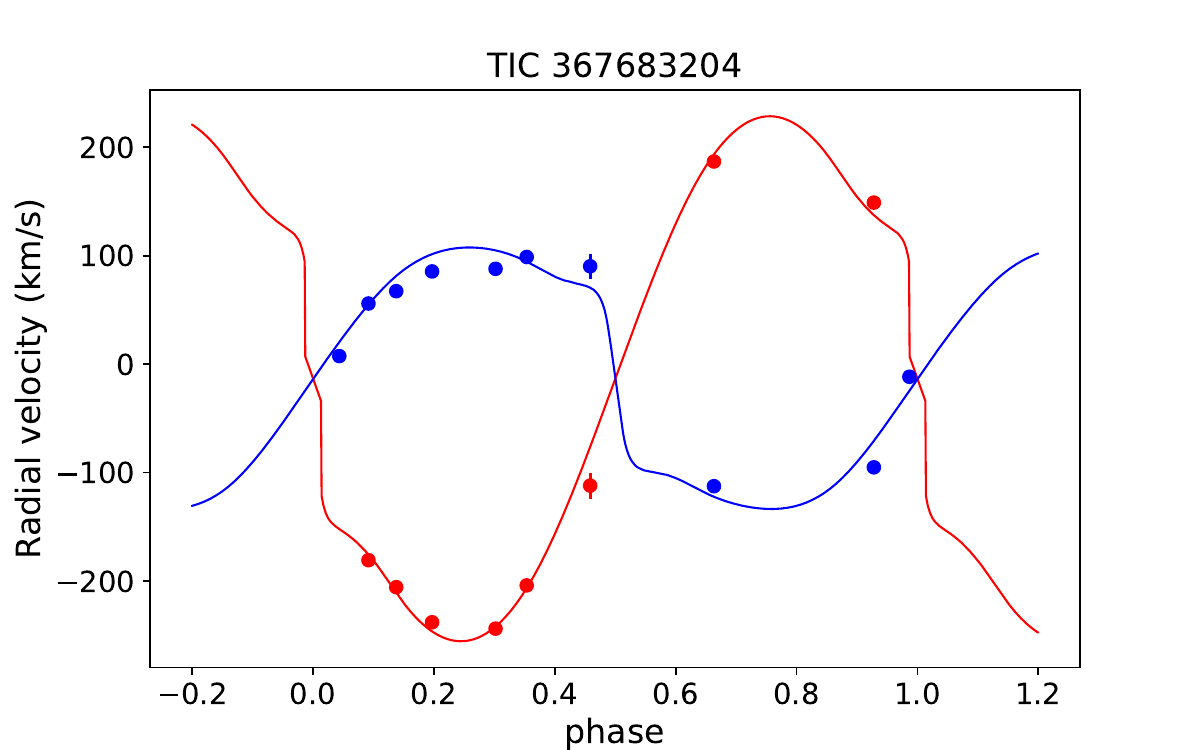}
    \end{minipage}
    
    \caption{These figures display the theoretical light curves and radial velocity curves for 11 targets. In the light curve fitting diagrams, the gray dots represent the observed data, and the red lines indicate the fitted theoretical curve. In the radial velocity diagrams, blue dots and lines denotes the more massive primary component, while red ones signifies the less massive secondary component.}
    \label{fig:lc}
\end{figure*}

We used the \emph{lightkurve} package \citep{2018ascl.soft12013L} to download TESS data with a 2-minute cadence \citep{https://doi.org/10.17909/fwdt-2x66} and selected the sector closest in time to the LAMOST spectra for analysis of the light curves. The sectors used for each target is listed in Table \ref{tab:basic information}. The downloaded data were phase-folded and binned into 1000 data points. The light curves are shown in Figure \ref{fig:lc}.

\subsection{Radial velocity measurements}

To measure radial velocities, only the blue-arm spectra were used to avoid the influence of potential H$\alpha$ emission lines. PHOENIX spectra were employed as template spectra. The cross-correlation function (CCF) method \citep{2007A&A...465..943S,2010AJ....140..184M} was used to calculate the CCF between the template spectra and the observed spectra. The GaussPy+ package \citep{2019A&A...628A..78R} was utilized to determine the peak positions of the CCF curves. The radial velocity values are listed in Table \ref{tab:RV} and the radial velocity curves of 11 targets are displayed in Figure \ref{fig:lc}.



\section{photometric solutions}

\begin{table}
\centering
\caption{The physical parameters of the 11 contact binaries.}
\label{tab:solution}
\begin{tabular}{ccccccccccccccc}
\hline
\hline
Targets(TIC) & $q$  & $T_{1}(K)$  &  $T_{2}(K)$  & $i(deg)$ &  $a(R_{\odot})$  &  $V_{\gamma}(km/s)$  &$f(\%)$ &    \\
\hline
    16617827  & 0.622$\pm$0.011  & 4958$\pm$2     & 5315$\pm$146     & 81.52$\pm$0.06  & 1.96$\pm$0.06  & -13.70$\pm$3.71 &  11.3$\pm$0.1 \\
    20212631  & 0.140$\pm$0.016  & 5420$\pm$3     & 5680$\pm$101     & 76.78$\pm$0.07  & 1.96$\pm$0.03  & -56.06$\pm$1.43 &  23.8$\pm$0.3 \\
    116484192 & 0.376$\pm$0.009  & 5363$\pm$4     & 5620$\pm$89     & 65.15$\pm$0.04  & 2.01$\pm$0.05  & -58.16$\pm$2.76 &  13.6$\pm$0.1 \\
    122712115 & 0.231$\pm$0.128  & 7989$\pm$87    & 7580$\pm$41     & 28.31$\pm$0.37  & 4.99$\pm$0.18  & -27.04$\pm$1.17 &  45.6$\pm$1.5 \\
    142587827 & 0.159$\pm$0.001  & 6700$\pm$146     & 6566$\pm$3     & 86.19$\pm$0.05  & 2.72$\pm$0.10  &   4.75$\pm$2.71 &  53.2$\pm$0.1 \\
    155968973 & 0.439$\pm$0.012  & 4573$\pm$16    & 5330$\pm$133     & 51.04$\pm$0.05  & 2.06$\pm$0.08  & -13.15$\pm$3.83 &  14.6$\pm$0.2 \\
    198410119 & 0.354$\pm$0.031  & 4965$\pm$575   & 5720$\pm$146     & 58.20$\pm$11.20 & 1.82$\pm$0.22  & -10.50$\pm$0.23 &  68.4$\pm$5.6 \\
    219109908 & 0.176$\pm$0.001  & 5840$\pm$142     & 5850$\pm$5     & 80.57$\pm$0.17  & 2.67$\pm$0.04  & -48.17$\pm$2.68 &  29.1$\pm$0.1 \\
    286169068 & 0.272$\pm$0.002  & 7130$\pm$104     & 7148$\pm$7     & 70.26$\pm$0.08  & 2.71$\pm$0.04  &  16.45$\pm$1.84 &  32.6$\pm$0.1 \\
    349294422 & 0.333$\pm$0.004  & 5620$\pm$100     & 5330$\pm$12    & 34.24$\pm$0.12  & 2.59$\pm$0.08  & -11.18$\pm$2.38 &  93.5$\pm$0.1 \\
    \vspace{1mm}
    367683204 & 0.518$\pm$0.003  & 3980$\pm$1     & 4160$\pm$14     & 85.67$\pm$0.06  & 1.66$\pm$0.04  & -13.49$\pm$3.64 &  13.9$\pm$0.1 \\  
\hline
Targets(TIC) & $M_{1}$($M_{\odot}$)  & $M_{2}$($M_{\odot}$)  & $R_{1}$($R_{\odot}$) & $R_{2}$($R_{\odot}$)  & $L_{1}$($L_{\odot}$)  & $L_{2}$($L_{\odot}$)  & $log J_{orb}($cgs$) $  \\
\hline
16617827  &   0.74$\pm$0.07 & 0.46$\pm$0.05    & 0.84$\pm$0.03   & 0.68$\pm$0.02   & 0.39$\pm$0.03     & 0.33$\pm$0.02   & 51.418  \\  
20212631  &   1.05$\pm$0.06 & 0.15$\pm$0.01    & 1.10$\pm$0.02   & 0.46$\pm$0.01   & 0.94$\pm$0.03     & 0.20$\pm$0.01   & 51.077  \\  
116484192 &   1.04$\pm$0.07 & 0.39$\pm$0.02    & 0.96$\pm$0.03   & 0.62$\pm$0.01   & 0.41$\pm$0.04     & 0.26$\pm$0.01   & 51.208  \\  
122712115 &   2.53$\pm$0.53 & 0.59$\pm$0.20    & 2.66$\pm$0.14   & 1.43$\pm$0.07   &25.88$\pm$3.74     & 6.05$\pm$0.59   & 52.055  \\  
142587827 &   1.21$\pm$0.14 & 0.19$\pm$0.02    & 1.53$\pm$0.06   & 0.71$\pm$0.03   & 4.25$\pm$0.33     & 0.83$\pm$0.07   & 51.290  \\  
155968973 &   0.70$\pm$0.08 & 0.31$\pm$0.04    & 0.95$\pm$0.04   & 0.66$\pm$0.02   & 0.54$\pm$0.03     & 0.18$\pm$0.02   & 51.173 \\   
198410119 &   0.69$\pm$0.27 & 0.25$\pm$0.10    & 0.94$\pm$0.22   & 0.62$\pm$0.23   & 0.48$\pm$0.45     & 0.37$\pm$0.28   & 51.156  \\  
219109908 &   1.24$\pm$0.05 & 0.22$\pm$0.01    & 1.53$\pm$0.02   & 0.76$\pm$0.02   & 2.22$\pm$0.07     & 0.53$\pm$0.03   & 51.329  \\  
286169068 &   1.22$\pm$0.06 & 0.33$\pm$0.02    & 1.39$\pm$0.02   & 0.79$\pm$0.02   & 4.50$\pm$0.14     & 1.47$\pm$0.09   & 51.511  \\  
349294422 &   1.82$\pm$0.17 & 0.61$\pm$0.06    & 1.40$\pm$0.05   & 0.92$\pm$0.07   & 1.75$\pm$0.12     & 0.62$\pm$0.10   & 51.840 \\   
367683204 &   0.83$\pm$0.07 & 0.43$\pm$0.04    & 0.74$\pm$0.02   & 0.55$\pm$0.01   & 0.12$\pm$0.01     & 0.08$\pm$0.00   & 51.391  \\  
\hline

\multicolumn{1}{c}{\multirow{5}*{star-spot}}  & \multicolumn{1}{c}{\multirow{2}*{Targets(TIC)} }  & \multicolumn{3}{c}{star1} &  \multicolumn{3}{c}{star2} \\

\cmidrule(r){3-5} \cmidrule(r){6-8}
& & $\lambda(deg)$   & $r_{s}(deg)$ &   $T_{s}$  &  $\lambda(deg)$   & $r_{s}(deg)$ &   $T_{s}$ \\
\cline{2-8}
& 20212631 	&   76 	  &  11 	&  0.726 	&  ...  &  ...  &	...     \\
& 198410119 &   ... 	&  ... 	&  ... 	  &  359 	&  26 	&  0.731  \\
& 367683204 &   ... 	&  ... 	&  ... 	  &  266 	&  10 	&  0.718  \\
\hline
\end{tabular}
\end{table}

The Wilson–Devinney (W-D) program \citep{1971ApJ...166..605W,1979ApJ...234.1054W,1990ApJ...356..613W} was used to analyze photometric light curves and radial velocity curves of the 11 stars. 
The temperatures of primary stars were adopted the mean temperatures from Gaia, TESS, and LAMOST surveys; 
the standard deviation of the sample of three temperatures is taken as the standard error of the mean. For targets with only one temperature data source, the standard deviation cannot be calculated; therefore, the maximum value of the standard deviations of the remaining targets is adopted as the error.
Mode 3 was applied to all systems. In this process, the orbital inclination, the temperature of the secondary star, the semi-major axis, the systemic radial velocity, the surface potential of the primary, the luminosity of the primary star and the third light were set as adjustable parameters.
The gravity-darkening coefficients $g_{1}$, $g_{2}$ and the bolometric albedos $A_{1}$, $A_{2}$ were determined according to \cite{1967ZA.....65...89L} and \cite{1969AcA....19..245R}, respectively. 
For the components of contact binaries with temperatures above and below 7200K, the gravity-darkening coefficients were set as 1.0 and 0.32, respectively \citep{1924MNRAS..84..665V, 1967ZA.....65...89L}, and the bolometric albedos were set to 1.0 and 0.5, respectively \citep{1969AcA....19..245R}.
\cite{1993AJ....106.2096V}'s table and van Hamme's peronal website\footnote[2]{https://faculty.fiu.edu/~vanhamme/lcdc2015/} updated in 2019 were adopted to determine the bolometric and bandpass limb-darkening coefficients.
For targets that exhibit a significant O'Connell effect, we have used a star-spot model and fixed the latitude of the star-spot at 90 degrees. 
The physical parameters are listed in Table \ref{tab:solution}. Only one target, TIC 198410119, exhibits a third light with $L_3/L_T=0.46$. This is likely attributed to the relatively large pixel scale of TESS, which results in contamination from numerous surrounding stars contributing to the third light.
It should be stated here that since the mass ratio refers to the ratio of the less massive component to the more massive one, the more massive component is defined as the primary star and the less massive one as the secondary star. This definition shall apply to all subsequent references to the primary and secondary stars.

\section{O-C Analysis}

\begin{table}
\caption{The minima times of 11 targets.}
\hspace*{-2cm} 
\scalebox{0.80}{
\label{tab:mini}
\begin{tabular}{cccccccccccccc}
\hline
\hline
TIC 16617827 & BJD & Error  & Epoch  & \textit{O-C} & residuals  & Source  & BJD & Error  & Epoch  & \textit{O-C} & residuals  & Source  \\
\hline  
&  2449999.30198 &  0.0003	&  -32415	  &  -0.00588 	& -0.00121 & (1)	&	 2453187.21901 &	0.0001  &    -21446	  &   -0.00191 	& 0.00212  & (1)	\\
&  2449999.44698 &  0.0002	&  -32414.5	&  -0.00620 	& -0.00152 & (1)	&	 2453234.29601 &	0.0010	&    -21284	  &   -0.00686 	& -0.00294 & (1)	\\
&  2450246.48297 &  0.0003	&  -31564.5	&  -0.00513 	& -0.00041 & (1)	&	 2453234.44201 &	0.0002	&    -21283.5	&   -0.00618 	& -0.00226 & (1)	\\
&  2450324.37197 &  0.0010	&  -31296.5	&  -0.00479 	& -0.00006 & (1)	&	 2453234.58701 &	0.0003	&    -21283	  &   -0.00649 	& -0.00257 & (1)	\\
&  2451770.39899 &  0.0003	&  -26321	  &  -0.00398 	& 0.00106  & (1)	&	 2453238.36501 &	0.0003	&    -21270	  &   -0.00668 	& -0.00276 & (1)	\\
&  2451770.54499 &  0.0003	&  -26320.5	&  -0.00329 	& 0.00175  & (1)	&	 2453238.51101 &	0.0003	&    -21269.5	&   -0.00599 	& -0.00208 & (1)	\\
&  2452203.29500 &  0.0005	&  -24831.5	&  -0.00034 	& 0.00469  & (1)	&	 2453250.28101 &	0.0003	&    -21229	  &   -0.00648 	& -0.00260 & (1)	\\
&  2452609.59201 &  0.0009	&  -23433.5	&  -0.00313 	& 0.00173  & (1)	&	 2453250.42801 &	0.0001	&    -21228.5	&   -0.00479 	& 0.00109  & (1)	\\
&  2452801.55301 &  0.0010	&  -22773	  &  -0.00280 	& 0.00189  & (1)	&	 2453250.43001 &	0.0002	&    -21228.5	&   -0.00279 	& -0.00091 & (1)	\\
&  2452835.41101 &  0.0001	&  -22656.5	&  -0.00312 	& 0.00154  & (1)	&	 2453250.57201 &	0.0002	&    -21228	  &   -0.00611 	& -0.00023 & (1)	\\
&  2452846.45401 &  0.0005	&  -22618.5	&  -0.00403 	& 0.00061  & (1)	&	 2453250.57401 &	0.0001	&    -21228	  &   -0.00411 	& -0.00223 & (1)	\\
&  2452859.53301 &  0.0015	&  -22573.5	&  -0.00335 	& 0.00127  & (1)	&	 2453251.29901 &	0.0002	&    -21225.5	&   -0.00568 	& -0.00180 & (1)	\\
&  2452897.46101 &  0.0001	&  -22443	  &  -0.00248 	& 0.00209  & (1)	&	 2453251.44401 &	0.0002	&    -21225	  &   -0.00600 	& -0.00212 & (1)	\\
&  2452981.30801 &  0.0043	&  -22154.5	&  -0.00204 	& 0.00240  & (1)	&	 2453251.58901 &	0.0003	&    -21224.5	&   -0.00631 	& -0.00243 & (1)	\\
&  2453168.76301 &  0.0003	&  -21509.5	&  -0.00295 	& 0.00113  & (1)	&	 2453258.42201 &	0.0013	&    -21201	  &   -0.00310 	& 0.00076  & (1)	\\
\hline
\end{tabular}
}
\vspace{3pt}
\begin{tablenotes}[]  
\item \hspace{0.35cm} \parbox{17.2cm}{\textbf{Note--} All the collected and calculated eclipsing times have been uniformly converted to BJD. “Epoch” represents the cycle number. "O - C" and “Residuals” represent the observed minima minus calculated minima and the corresponding fitting residuals, respectively. “Source” represents the source of the observed minima. (1) VarAstro; (2) WASP; (3) ASAS-SN; (4) TESS; (5) AAVSO; (6) ZTF; This table is available in its entirety in machine-readable form. }
\end{tablenotes} 
\end{table}

The O-C (observed eclipsing minimum minus calculated eclipsing minimum) analysis is a powerful probe of how eclipsing binaries evolve and whether they host companions \citep{2016ApJ...817..133Z, 2019ApJ...871L..10E, 2019RAA....19..147L, 2019AJ....157..207L, 2019ApJ...877...75P, 2025PASP..137h4201P, 2025AJ....170..214P}. 
We have assembled all available eclipsing times for these 10 systems, drawing from Wide Angle Search for Planets \citetext{SuperWASP, \citealp{2010A&A...520L..10B}}, All-Sky Automated Survey for SuperNovae \citetext{ASAS-SN, \citealp{2014ApJ...788...48S, 2018MNRAS.477.3145J}}, VarAstro\footnote[3]{VarAstro (https://var.astro.cz/en/).} and AAVSO\footnote[4]{AAVSO (https://www.aavso.org/).} and TESS. 

\cite{2024ApJ...961...97Z} has conducted a detailed analysis of the orbital period changes for TIC 20212631. The data used in the analysis are almost identical to ours. Therefore, we no longer repeat the analysis of the orbital period changes of TIC 20212631.
For the remaining 10 targets, we preferentially adopted the eclipsing minima from high-quality CCD and photoelectric observations in VarAstro; for published data lacking uncertainties, we assigned a uniform error of 0.001.
All the eclipsing times have been unified and converted to BJD using the online tool\footnote[5]{https://astroutils.astronomy.osu.edu/time/hjd2bjd.html}, then a linear correction was applied to each target.
The O-C values were recalculated using the newly obtained P and $BJD_{0}$ with linear correction, through following equation:
\begin{equation}
BJD=BJD_{0}+P\times E,
\end{equation}
where  E refers to the number of cycles. The latest O-C values are listed in Table \ref{tab:mini} and the corresponding plots are in Figure \ref{fig:oc}. 
After applying a linear correction, the residuals of O-C for TIC 122712115 became flat without any discernible trend, hence only linear correction was applied to TIC 122712115. Among the remaining 9 targets, 6 exhibited parabolic trends (TIC 116484192, TIC 142587827, TIC 155968973, TIC 198410119,  TIC 286169068, TIC 349294422), while 3 targets showed parabolic plus periodic variations (TIC 16617827, TIC 367683204, TIC 219109908). 
For those exhibiting a parabolic trend, we used the following equation to fit their O-C curves:
\begin{equation}
O-C=\Delta T_{0}+ \Delta P_{0} \times E + \frac{\beta}{2} \times E^{2},
\end{equation}
where $\Delta T_{0}$, $\Delta P_{0}$ refer to the corrections of the initial epoch and orbital period, respectively, $\beta$ refers to the long-term changing rate of the orbital period.
For targets with parabolic plus periodic variations in the O-C curves, we employed two models for fitting, one of which is the model of a parabola plus a sine function, as shown in the following equation:
\begin{equation}
O-C=\Delta T_{0}+ \Delta P_{0} \times E + \frac{\beta}{2} \times E^{2} + A \times sin (\frac{2 \pi}{P_{3}} \times E + \phi),
\label{sin}
\end{equation}
where A, $P_{3}$, and $\phi$ represent the amplitude, period, and initial phase of the sinusoidal change. 
Another model involves orbital period changes caused by the light-travel time effect (LTTE; \citealp{1922BAN.....1...93W}; \citealp{1959AJ.....64..149I}) due to the third body, which can be fitted using the following equation:
\begin{equation}
O-C=\Delta T_{0}+ \Delta P_{0} \times E + \frac{\beta}{2} \times E^{2} + A [{(1-e^{2})} \frac{\sin (\nu+\omega)}{(1+e \cos \nu)}+e \sin \omega],
\end{equation}
where $e$, $\omega$, $\nu$ are the eccentricity, longitude of periastron and true anomaly of the third body's orbit, respectively \citep{1952ApJ...116..211I}.
From the above equation, it is evident that when \( e=0 \) , Equation (4) reduces to Equation (3).
The fitting parameters of the O-C curves are listed in Table \ref{tab:oc} and the O-C fitting curves are plotted in Figure \ref{fig:oc} for the 10 targets (except for TIC 20212631).

For the 3 targets exhibiting periodic variations, 
we fitted the O-C curves of two of them TIC 16617827 and TIC 367683204 using a parabola superimposed on the LTTE model. Regarding TIC 219109908, it is worth noting that a single function is difficult to adapt to the fitting requirements of these two segments of data. Therefore, we fitted each segment separately. We fitted the two segments of TIC 219109908 separately using Equation (3), where the amplitude and period of the sinusoidal component in the second segment were adopted from the fitting results of the first segment.
In the figure, “fitting curve 1” and “fitting curve 2” represent the different fitting curves for these two segments, respectively.

\section{Discussion and Conclusions}

Using the W-D program, we performed analyses of the light and radial velocity curves for 11 TESS contact binaries and derived their physical parameters. Among the 11 targets, there are 5 shallow-contact binaries, 3 moderate-contact binaries, and 3 deep-contact binaries. There are 3 targets exhibiting O’Connell effect, which is interpreted using one star-spot on the primary component. Among these targets, 
TIC 16617827, TIC 20212631, TIC 122712115, TIC 142587827, TIC 155968973, TIC 349294422, and TIC 367683204 have been previously studied. Our results are nearly consistent with the historical findings, with differences in mass ratio within 3\%—except for TIC 20212631, the mass ratio from \cite{2011AJ....142...99C} is 0.207, which shows a larger discrepancy. This may be attributed to the fact that the historical results relied solely on photometric data, leading to greater errors.

\subsection{Orbital Period Variations}
\subsubsection{Secular Period Increase or Decrease}
The periods of TIC 116484192, TIC 198410119, TIC 349294422, TIC 16617827 and TIC 367683204 are decreasing in the long term. The long-term period decrease may be caused by mass transfer from the more massive component to the less massive one, angular momentum loss, or a combination of both. To verify which is responsible for the period decrease, we assume that the period decrease is due to angular momentum loss. The rate of period decrease caused by angular momentum loss can be calculated using the following equation \citep{1967ApJ...148..217W, 1988ASIC..241..345G}:
\begin{equation}
\dot{P}_{aml}=1.1 \times 10^8 q^{-1}(1+q)^2(M_{1}+M_{2})^{-5/3}k^2\times (M_{1}R_{1}^{4}+M_{2}R_{2}^{4})P^{-7/3},
\end{equation}
where \( k \) is the dimensionless gyration radius. The value of \( k^2 = 0.06 \) was adopted from \citet{2006MNRAS.369.2001L}. 
The unit of the leading constant is d \( yr^{-1} \). $M_{1}$, $M_{2}$, $R_{1}$ and $R_{2}$ are in solar units, and P is in days.
The calculated rates of decrease are \( 4.31 \times 10^{-8} \), \( 3.83 \times 10^{-8} \), \( 9.07 \times 10^{-8} \), \( 1.76 \times 10^{-8} \) and \( 2.01 \times 10^{-8} \) d \( yr^{-1} \), all of which are smaller than the observed period decrease rates of the three targets. Therefore, the period decrease is probably caused by the combination of mass transfer and angular momentum loss. The mass transfer rate can be calculated using the following equation:
\begin{equation}
\dot{M}_{1}=-\dot{M}_{2}=\frac{M_{1}M_{2}}{3P(M_{1}-M_{2})}\times \dot{P}.
\end{equation}
The values are listed in Table \ref{tab:oc}. The periods of the targets TIC 142587827, TIC 155968973, TIC 286169068 and TIC 219109908 are increasing in the long term. The cause of the period increase is generally attributed to mass transfer from the less massive component to the more massive one. 
The mass transfer rates are listed in Table \ref{tab:oc}.

\begin{table}[h]
\caption{The fitting parameters O-C analysis and mass transfer parameters of 10 targets.}
\label{tab:oc}
\begin{tabular}{cccccccccccc}
\hline
\hline
Targets(TIC) & $\Delta T_{0}$ (days) & $\Delta P_{0}$ (days) & $\beta$ (days $yr^{-1}$) & $dM_{1}/dt$ (M$_{\odot}$ $yr^{-1}$) \\
 & $\times 10^{-4}$ & $\times 10^{-6}$ & $\times 10^{-7}$ & $\times 10^{-7}$ \\
\hline
16617827        &  $-182.37 (\pm 1.70)  $    &    $-1.62 (\pm 0.05)$    &    $-0.93 (\pm 0.04)$   &    $-1.30(\pm0.59) $  \\  
116484192       &  $-3.75 (\pm 0.45)    $    &    $-0.30 (\pm 0.02)$    &    $-0.99 (\pm 0.05)$   &    $-0.40(\pm0.07)  $  \\ 
122712115       &   \dots                    &      \dots               &    \dots                &    \dots      \\          
142587827       &  $17.62 (\pm 0.32)    $    &    $2.86 (\pm 0.02) $    &    $2.47 (\pm 0.05) $   &    $0.43(\pm0.10)   $  \\ 
155968973       &  $1.15 (\pm 1.95)     $    &    $0.08 (\pm 0.08) $    &    $0.16 (\pm 0.13) $   &    $0.07(\pm0.06)   $  \\ 
198410119       &  $35.86 (\pm 0.51)    $    &    $-0.67 (\pm 0.01) $   &    $-1.86 (\pm 0.02)$   &    $-0.80(\pm0.81)  $  \\ 
$219109908^{1}$ &  $710.01 (\pm 192.77 )$    &    $15.83 (\pm 1.73) $   &    $7.77 (\pm 0.49) $   &    $1.58(\pm0.17)  $  \\  
$219109908^{2}$ &  $95.21 (\pm 68.37 )$      &    $-2.42 (\pm 1.04) $   &    $4.36 (\pm 2.18) $   &    $0.89(\pm0.45)  $  \\  
286169068       &  $-1.90 (\pm 0.30)    $    &    $-0.20(\pm 0.02) $    &    $1.72 (\pm 0.07) $   &    $0.63(\pm0.08)  $  \\  
349294422       &  $6.59 (\pm 0.38)     $    &    $0.45 (\pm 0.02) $    &    $-6.70 (\pm 0.16)$   &    $-6.54(\pm1.53) $  \\  
367683204       &  $-278.55 (\pm 26.56) $   &    $-1.98(\pm 0.23) $    &    $-0.72(\pm0.08)$   &    $-0.97(\pm2.07) $  \\  

\hline
\end{tabular}
\begin{tablenotes}[]  
\item \vspace{0.1cm}\hspace{2.95cm} \parbox{25.2cm}{\textbf{Note--} \newline $^{1}$Parameters obtained from fitting curve 1 of TIC 219109908. \newline $^{2}$Parameters obtained from fitting curve 2 of TIC 219109908. 
}
\end{tablenotes} 
\end{table}

\begin{table}
\hspace*{-1cm} 
\begin{threeparttable}
\caption{Parameters determined by analyzing the orbital period change of 3 targets} 
\label{tab:applegate}
\begin{tabular}{lcccccccc} 
\hline
\hline
Parameter & \multicolumn{2}{c}{TIC 219109908} & \multicolumn{2}{c}{TIC 16617827} & \multicolumn{2}{c}{TIC 367683204} & Unit \\
\cline{2-3} \cline{4-5} \cline{6-7}
 & Primary & Secondary & Primary & Secondary & Primary & Secondary & \\
\hline
$\Delta P$                     & 0.2851 & 0.2851 & 0.1276 & 0.1276 & 0.0938 & 0.0938 & s \\
$\Delta P/P$                   & $7.79\times10^{-6}$ & $7.79\times10^{-6}$ & $5.08\times10^{-6}$ & $5.08\times10^{-6}$ & $4.92\times10^{-6}$ & $4.92\times10^{-6}$ &  \\
$\Delta Q$                     & $7.19\times10^{49}$ & $1.27\times10^{49}$ & $1.55\times10^{49}$ & $0.96\times10^{49}$ & $1.20\times10^{49}$ & $6.24\times10^{48}$ & g cm$^2$ \\
$\Delta J$                     & $1.71\times10^{47}$ & $4.35\times10^{46}$ & $9.11\times10^{46}$ & $6.64\times10^{46}$ & $8.84\times10^{46}$ & $5.78\times10^{46}$ & g cm$^2$ s$^{-1}$ \\
$I_s$                          & $1.79\times10^{54}$ & $7.77\times10^{52}$ & $3.35\times10^{53}$ & $1.37\times10^{53}$ & $2.92\times10^{53}$ & $8.35\times10^{52}$ & g cm$^2$ \\
$\Delta\Omega$                 & $9.59\times10^{-8}$ & $5.59\times10^{-7}$ & $2.72\times10^{-7}$ & $4.86\times10^{-7}$ & $3.03\times10^{-7}$ & $6.92\times10^{-7}$ & s$^{-1}$ \\
$\Delta E$                     & $3.29\times10^{40}$ & $4.86\times10^{40}$ & $4.95\times10^{40}$ & $6.45\times10^{40}$ & $5.35\times10^{40}$ & $7.99\times10^{40}$ & erg \\
$\Delta L_{\mathrm{rms}}$      & $8.76\times10^{31}$ & $1.29\times10^{32}$ & $2.55\times10^{32}$ & $3.33\times10^{32}$ & $1.02\times10^{32}$ & $1.53\times10^{32}$ & erg s$^{-1}$ \\
$L_{\mathrm{rms}}/L_{\odot}$   & 0.0229 & 0.0338 & 0.0667 & 0.0869 & 0.0268 & 0.0400 & $L_{\odot}$ \\
$L_{\mathrm{rms}}/L_{1,2}$     & 0.0095 & 0.0567 & 0.1711 & 0.2634 & 0.2231 & 0.4998 & $L_{1,2}$ \\
$\Delta m$                     & $\pm0.0082$ & $\pm0.0122$ & $\pm0.0963$ & $\pm0.1238$ & $\pm0.1364$ & $\pm0.1979$ & mag \\
$B$                            & 4822 & 6936 & 11877 & 13917 & 8618 & 10874 & G \\
\hline
$A$                            &   \multicolumn{2}{c}{0.0170$\pm$0.0084}       & \multicolumn{2}{c}{0.0057$\pm$0.0003}         & \multicolumn{2}{c}{0.0149$\pm$0.0028}  &   $days$         \\  
$a \sin i $                    &   \multicolumn{2}{c}{2.93$\pm$1.45}           & \multicolumn{2}{c}{1.00$\pm$0.06}             & \multicolumn{2}{c}{2.58$\pm$0.49}      &   $AU$         \\  
$f(m) $                        &   \multicolumn{2}{c}{0.018$\pm$0.027}           & \multicolumn{2}{c}{0.002$\pm$0.001}          & \multicolumn{2}{c}{0.006$\pm$0.003}    &   $M_{\odot}$  \\    
$m_{3} $                       &   \multicolumn{2}{c}{0.40$\pm$0.38}           & \multicolumn{2}{c}{0.17$\pm$0.02}             & \multicolumn{2}{c}{0.24$\pm$0.09}      &   $M_{\odot}$  \\    
$a_{3}$                        &   \multicolumn{2}{c}{12.63$\pm$13.62}         & \multicolumn{2}{c}{8.27$\pm$1.03}             & \multicolumn{2}{c}{15.88$\pm$6.44}     &   $AU$  \\  
$e$                            &   \multicolumn{2}{c}{0}                       & \multicolumn{2}{c}{0.80$\pm$0.07}             & \multicolumn{2}{c}{0.27$\pm$0.03}      &          \\    
$\omega$                       &   \multicolumn{2}{c}{...}                     & \multicolumn{2}{c}{15.7$\pm$2.7}              & \multicolumn{2}{c}{38.2$\pm$12.2}      &   $^\circ$  \\    
$P_{3} $                       &   \multicolumn{2}{c}{37.42$\pm$8.04}          & \multicolumn{2}{c}{19.3$\pm$0.5}              & \multicolumn{2}{c}{52.0$\pm$4.8}       &   $yr$  \\     
$T_{3}$                        &   \multicolumn{2}{c}{...}                     & \multicolumn{2}{c}{2446878$\pm$180}        & \multicolumn{2}{c}{2456890$\pm$306}    &     \\
\hline
\end{tabular}
\end{threeparttable}
\end{table}

\begin{figure*}
    \centering
    \begin{minipage}[t]{0.48\textwidth}
        \centering
        \includegraphics[width=\linewidth, height=0.53\linewidth]{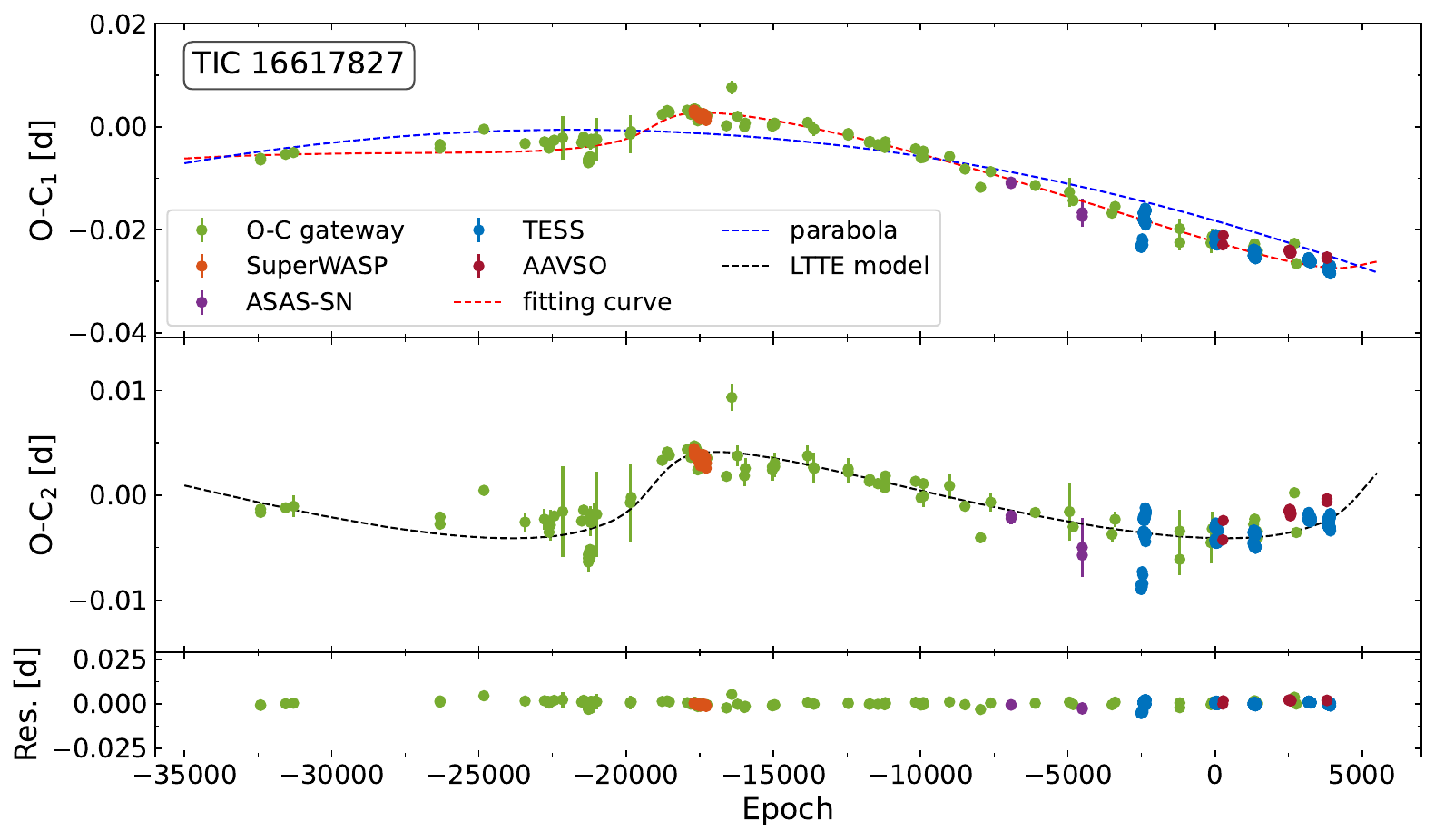}
    \end{minipage}
    \hfill
    \begin{minipage}[t]{0.48\textwidth}
        \centering
        \includegraphics[width=\linewidth, height=0.53\linewidth]{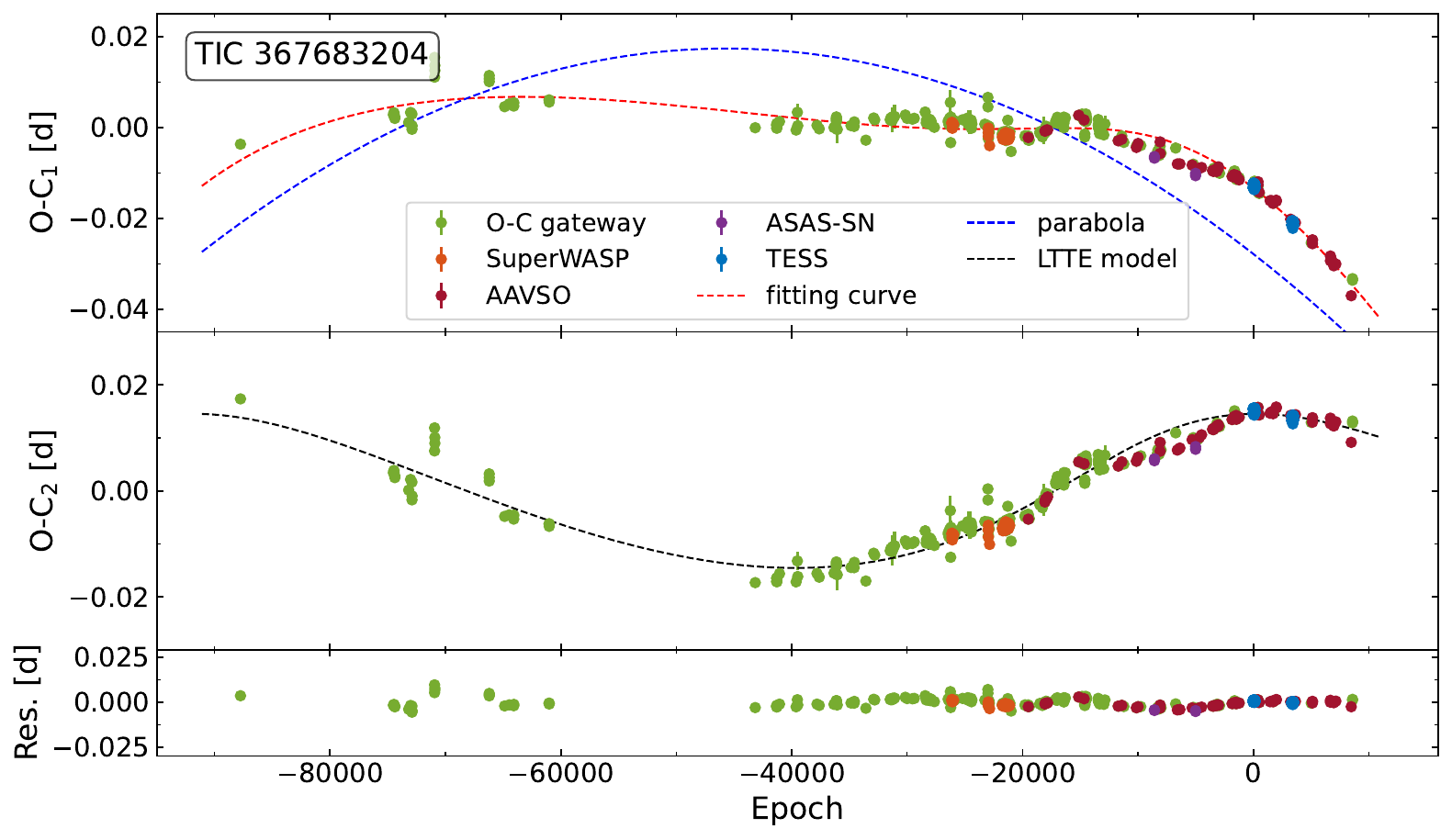}
    \end{minipage}
    
    \begin{minipage}[t]{0.48\textwidth}
        \centering
        \includegraphics[width=\linewidth, height=0.53\linewidth]{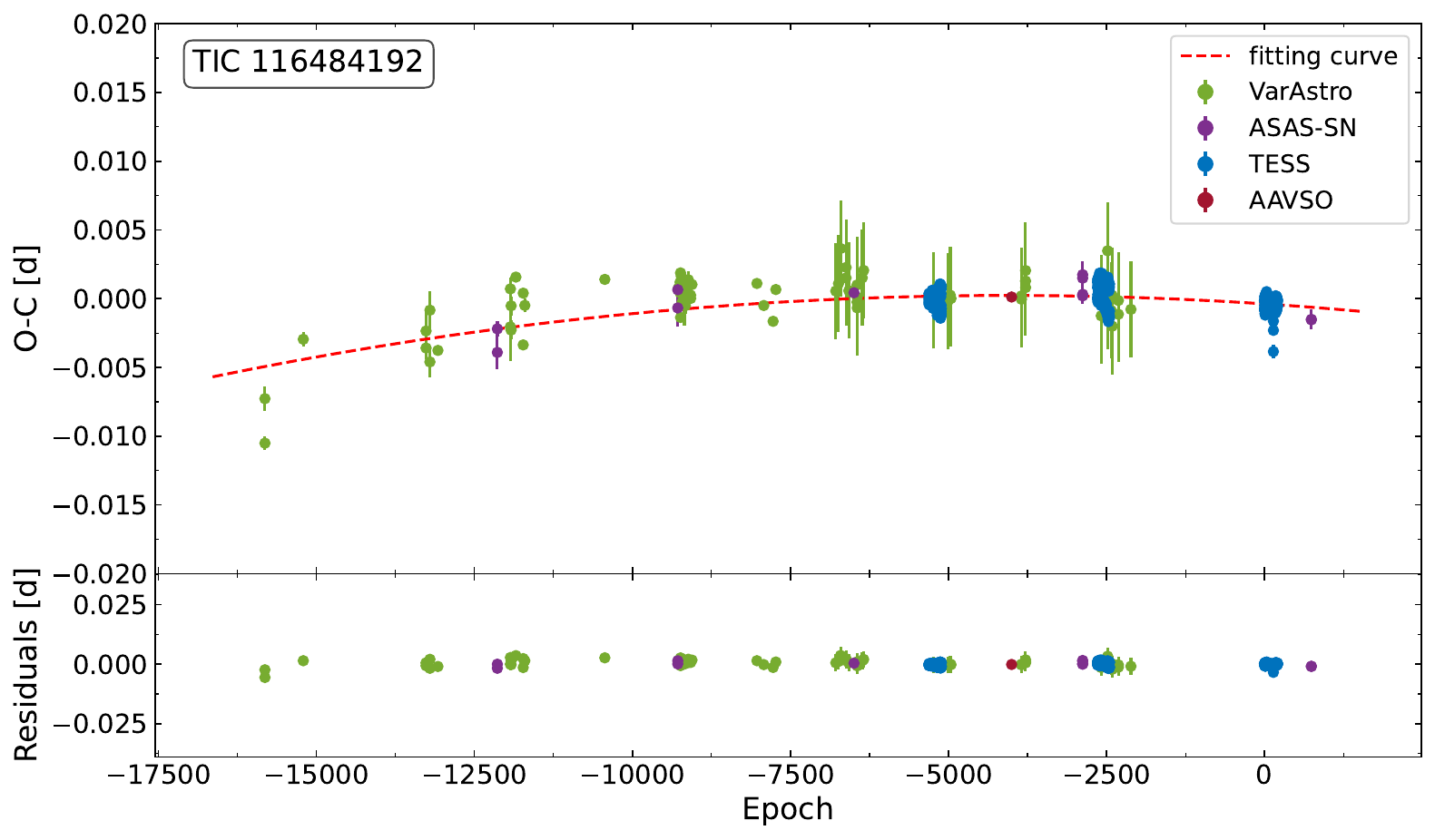}
    \end{minipage}
    \hfill
    \begin{minipage}[t]{0.48\textwidth}
        \centering
        \includegraphics[width=\linewidth, height=0.53\linewidth]{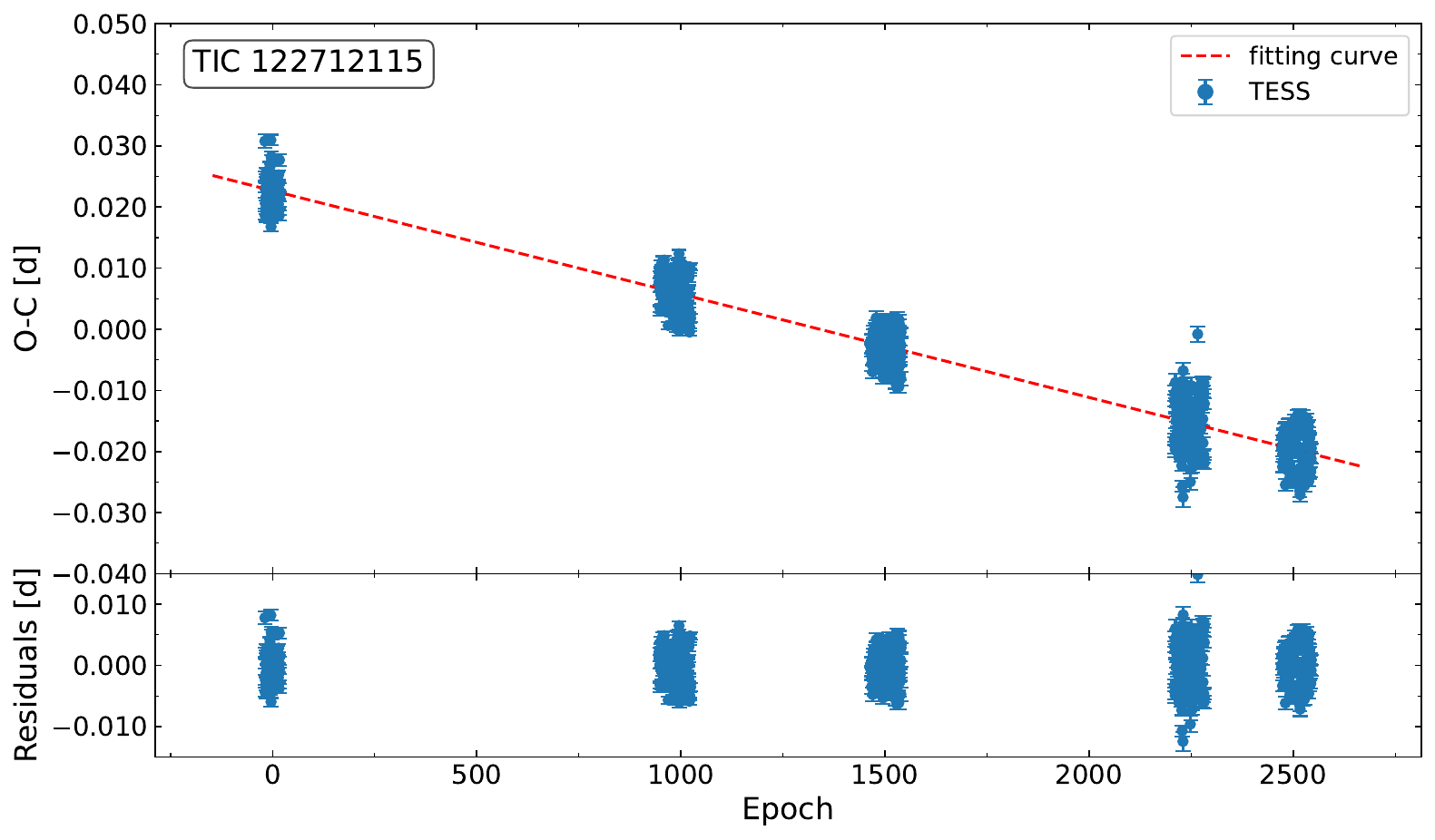}
    \end{minipage}
    
    \begin{minipage}[t]{0.48\textwidth}
        \centering
        \includegraphics[width=\linewidth, height=0.53\linewidth]{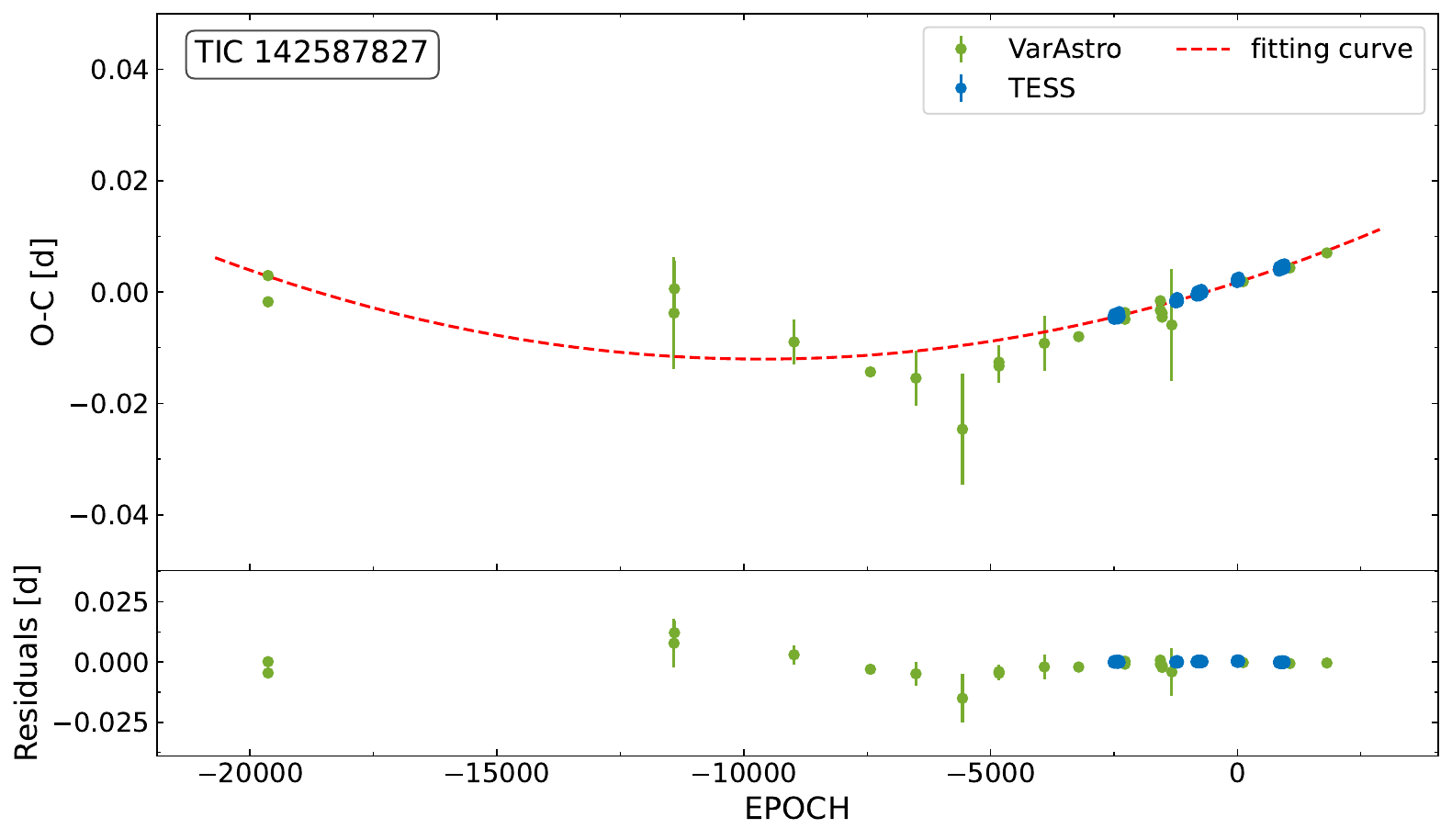}
    \end{minipage}
    \hfill
    \begin{minipage}[t]{0.48\textwidth}
        \centering
        \includegraphics[width=\linewidth, height=0.53\linewidth]{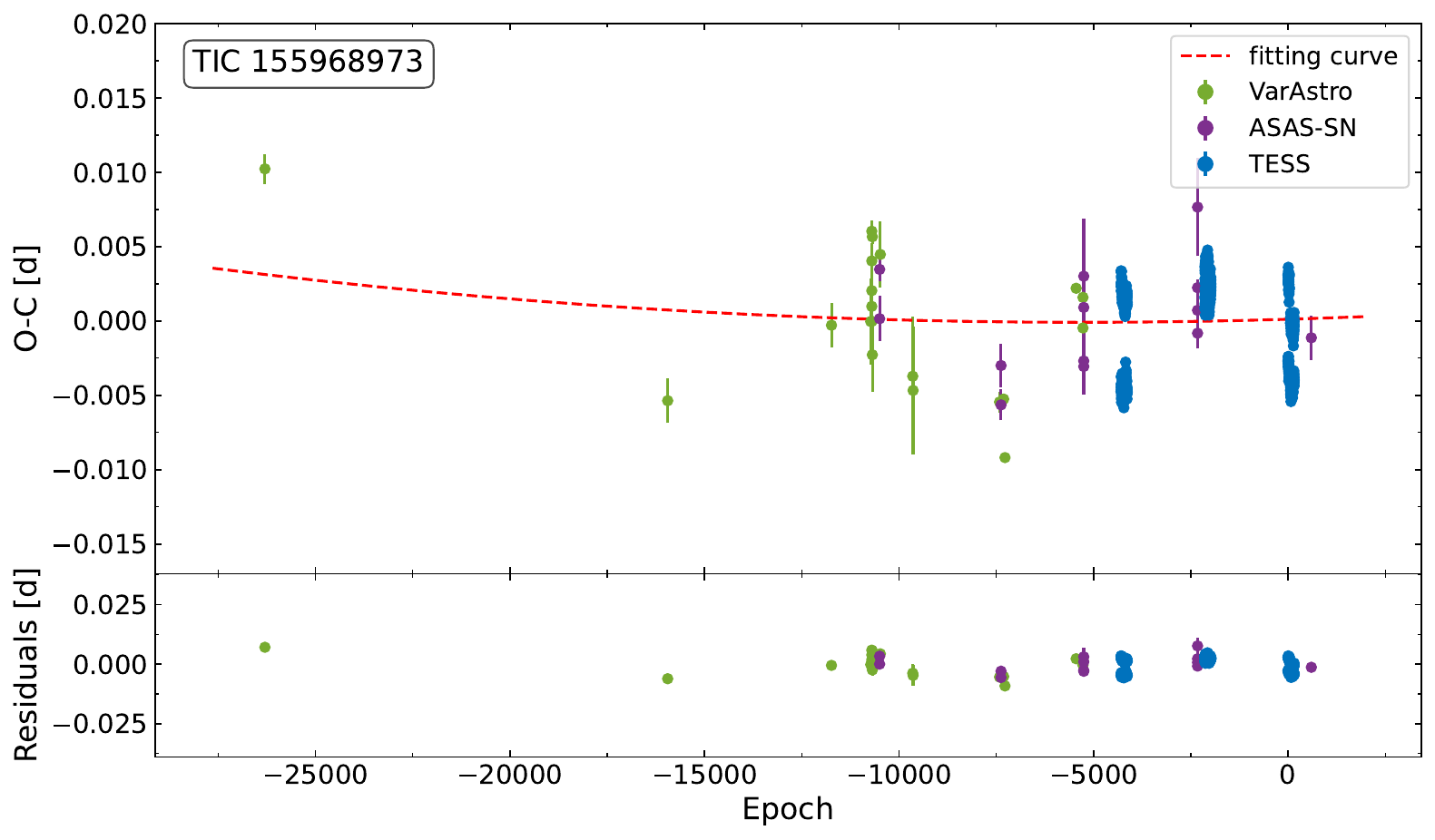}
    \end{minipage}
    
    \begin{minipage}[t]{0.48\textwidth}
        \centering
        \includegraphics[width=\linewidth, height=0.53\linewidth]{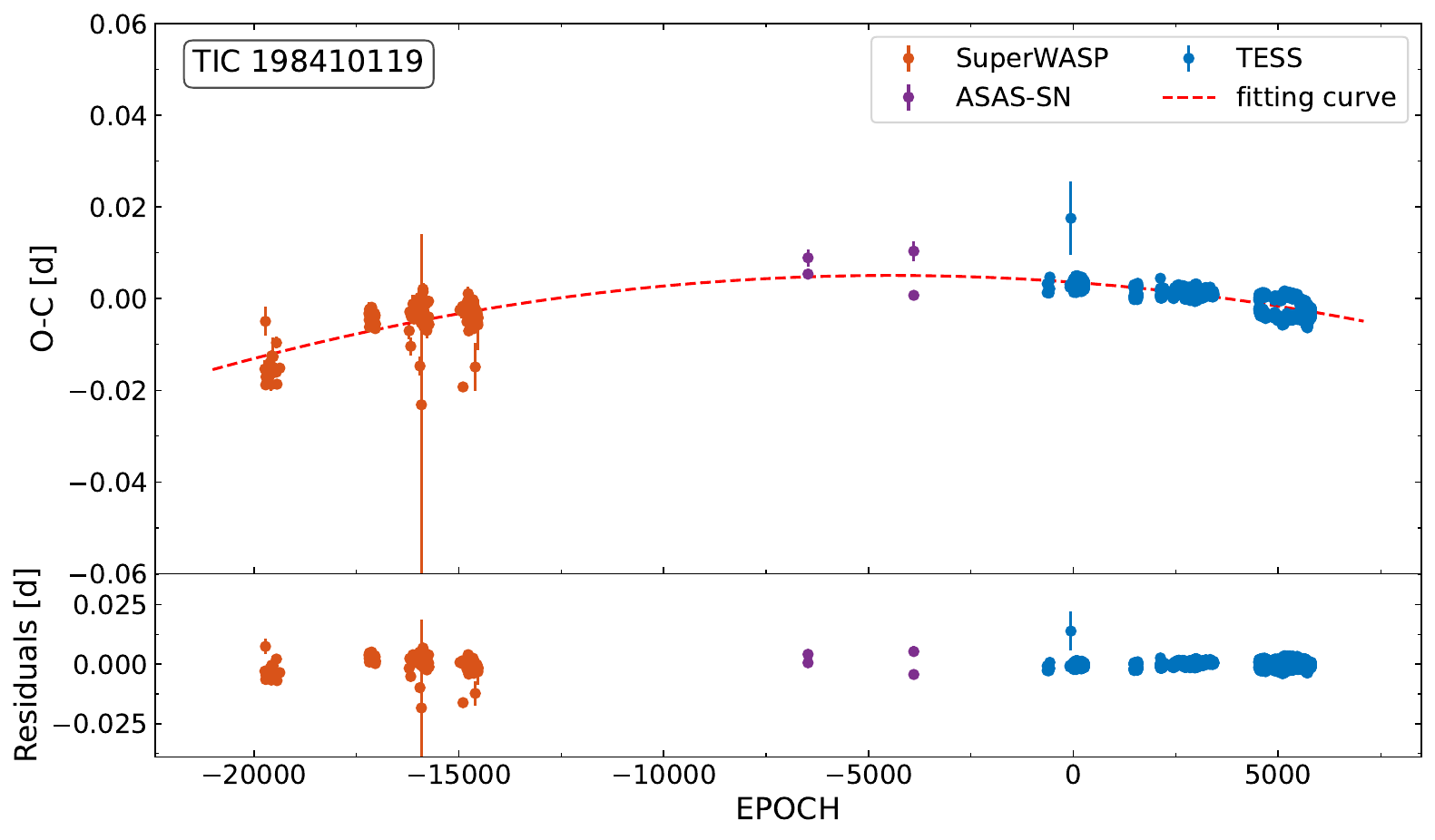}
    \end{minipage}
    \hfill
    \begin{minipage}[t]{0.48\textwidth}
        \centering
        \includegraphics[width=\linewidth, height=0.53\linewidth]{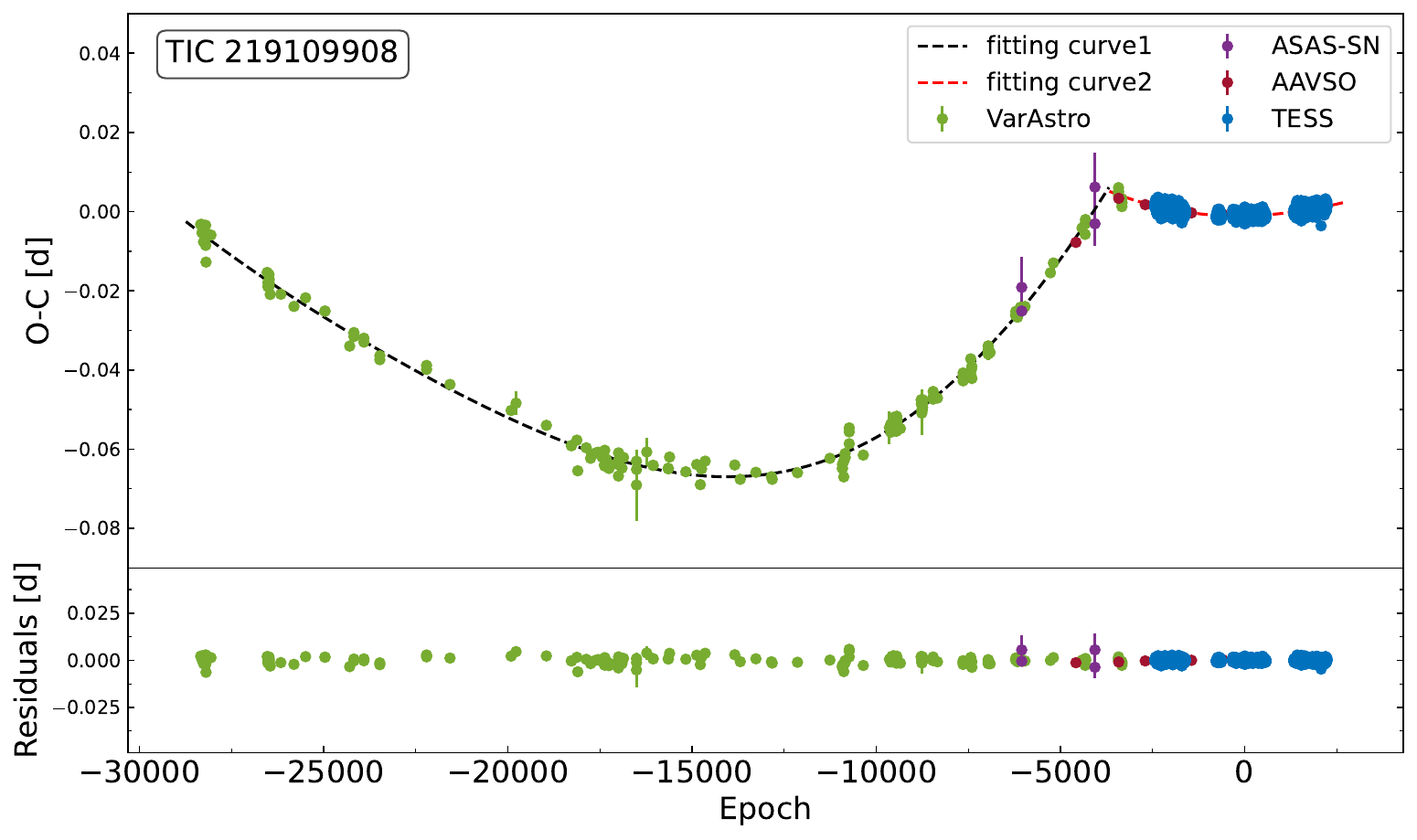}
    \end{minipage}
    
    \begin{minipage}[t]{0.48\textwidth}
        \centering
        \includegraphics[width=\linewidth, height=0.53\linewidth]{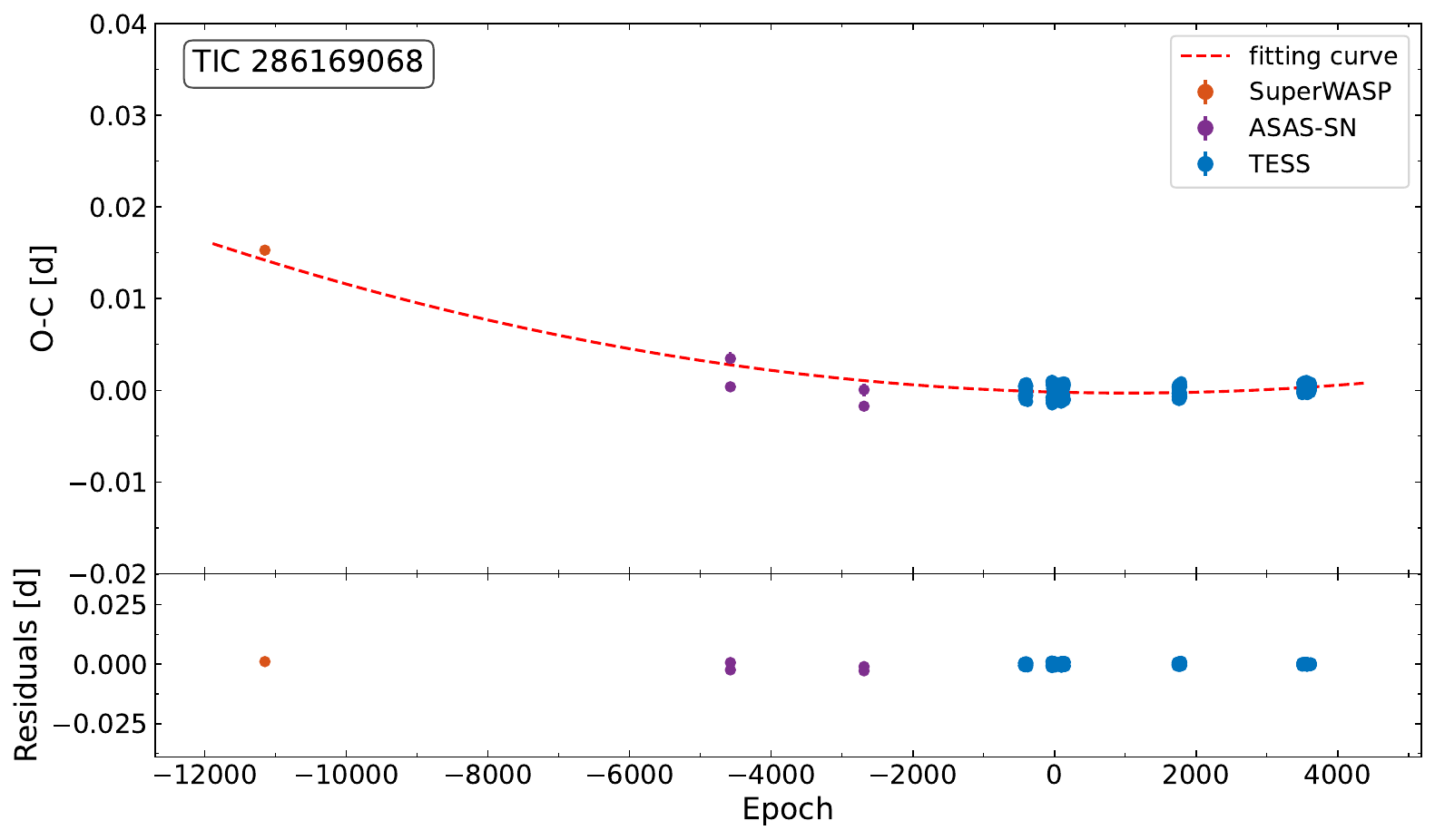}
    \end{minipage}
    \hfill
    \begin{minipage}[t]{0.48\textwidth}
        \centering
        \includegraphics[width=\linewidth, height=0.53\linewidth]{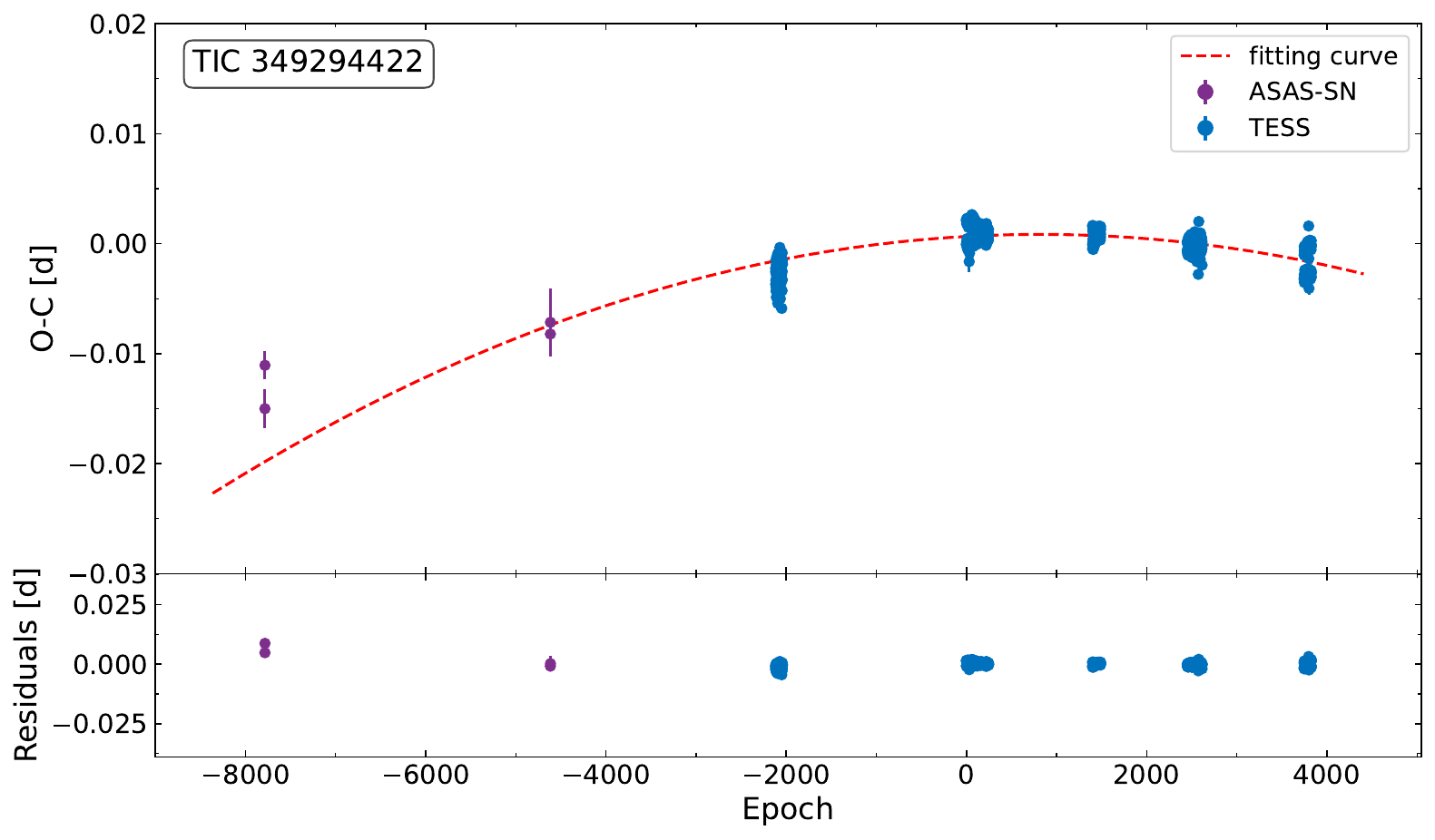}
    \end{minipage}
    \caption{The O-C values and fitting curves of 10 targets.}
    \label{fig:oc}
    \end{figure*}

    \clearpage

\subsubsection{Cyclic Period Variation}

Applegate's theory \citep{1992ApJ...385..621A} suggests that in binary systems, magnetic activity cycles can induce internal structural changes, which in turn drive periodic variations in the orbital period through tidal interactions with the orbit. If magnetic activity is responsible for the observed periodic variations in the orbital period, the required variation in the gravitational quadrupole moment, $\Delta Q$, can be determined using the following equation \citep{2002AN....323..424L}:
\begin{equation}
\frac{\Delta P}{P}=\frac{2\pi \times A}{P_{mod}},
\end{equation}
\begin{equation}
\frac{\Delta P}{P}= -9\frac{\Delta Q}{Ma^{2}},
\end{equation}
where, $\Delta P$/$P$ represents the amplitude of orbital period modulations, $P_{mod}$ refers to modulation period of the periodic variation in the orbital period, and $M$, $R$, and $a$ denote the mass, radius of the magnetically active component and the semimajor axis of the binary system, respectively. Based on this relationship, we can calculate $\Delta Q_{1}$ and $\Delta Q_{2}$ for the primary and secondary components. The values are listed in Table \ref{tab:applegate}. 
The transferred angular momentum and the energy required for this variation can be obtained using the following equation:
\begin{equation}
	\Delta J = -\frac{G M^2 }{R} (\frac{a}{R}^{2}) \frac{\Delta P}{6 \pi},
\end{equation}
\begin{equation}
	\Delta E = \Delta J \Delta \Omega + \frac{(\Delta J)^2}{2 I_{\text{eff}}}.
\end{equation}
The change in luminosity caused by the change in angular momentum can be calculated using the following equation:
\begin{equation}
	\Delta L_{\text{rms}} = \pi \frac{\Delta E}{ P_{\text{mod}}}.
\end{equation}
The luminosity variation converted to the magnitude scale is given by the following equation \citep{1997AJ....114.2753K}:
\begin{equation}
	\Delta m = -2.5 \log (1\pm\frac{\Delta L_{rms}}{ L_{p}+L_{s}}).
\end{equation}
The magnetic field required to produce these variations is as follows:
\begin{equation}
    B = (10 \frac{GM^2}{R^4}(\frac{a}{R})^2\frac{\Delta P}{P_{mod}})^{1/2},
\end{equation}
where the physical quantities are in the cgs units, with its leading constant being a dimensionless quantity.
These parameters of the Applegate model are listed in Table \ref{tab:applegate}. It can be found that the Applegate mechanism can lead to the periodic changes in the orbital period.

At the same time, we also considered the impact of the third body on the variation of the orbital period.
As can be seen from Figure \ref{fig:oc}, the O-C curve fitted using the sine fuction or LTTE model is very good for TIC 16617827, TIC 219109908 and TIC 367683204. Studies have shown that contact binary systems often have a companion star, and the presence of the companion star causes periodic changes in the orbital period of the system \citep{2012AJ....144..136Y, 2021ApJ...922..122L}. Therefore, it is highly likely that these three systems have companion stars. Assuming that the periodic changes in the orbital period of the system are caused by a companion star, according to the projected distance \( a_{12} \sin i_3 = A \times c \) between the eclipsing binary and the barycenter of the triple system, \( a_{12} \sin i_3 \) can be obtained, where \( i_3 \) is the inclination of the third body's orbit, \( c \) is the speed of light, and \( A \) is the amplitude. Then the mass function can be determined according to the following equation:
\begin{equation}
f(m_{3}) = \frac{ 4\pi^{2}}{GP_{3}^{2}} \times (a_{12}\sin i_{3})^{3} = \frac{(M_{3} \sin i_{3})^3}{(M_{1} + M_{2} + M_{3})^2}.
\end{equation}
The orbital distance between the third body and the inner binary can be estimated according to $a_{3}=(M_{1}+M_{2})\times a_{12}/M_{3}$. Assuming a coplanar orbit, the mass of the third body and orbital distance can be determined. The detail information is listed in Table \ref{tab:applegate}. From the table, TIC 16617827, TIC 219109908 and TIC 367683204 may have a red dwarf companion star.
To sum up, both the Applegate mechanism and the third body can potentially cause periodic changes in the orbital period.

\subsection{The Evolutionary States}

\begin{figure*}
    \begin{minipage}{0.55\textwidth}
        \centerline{\includegraphics[width=\linewidth]{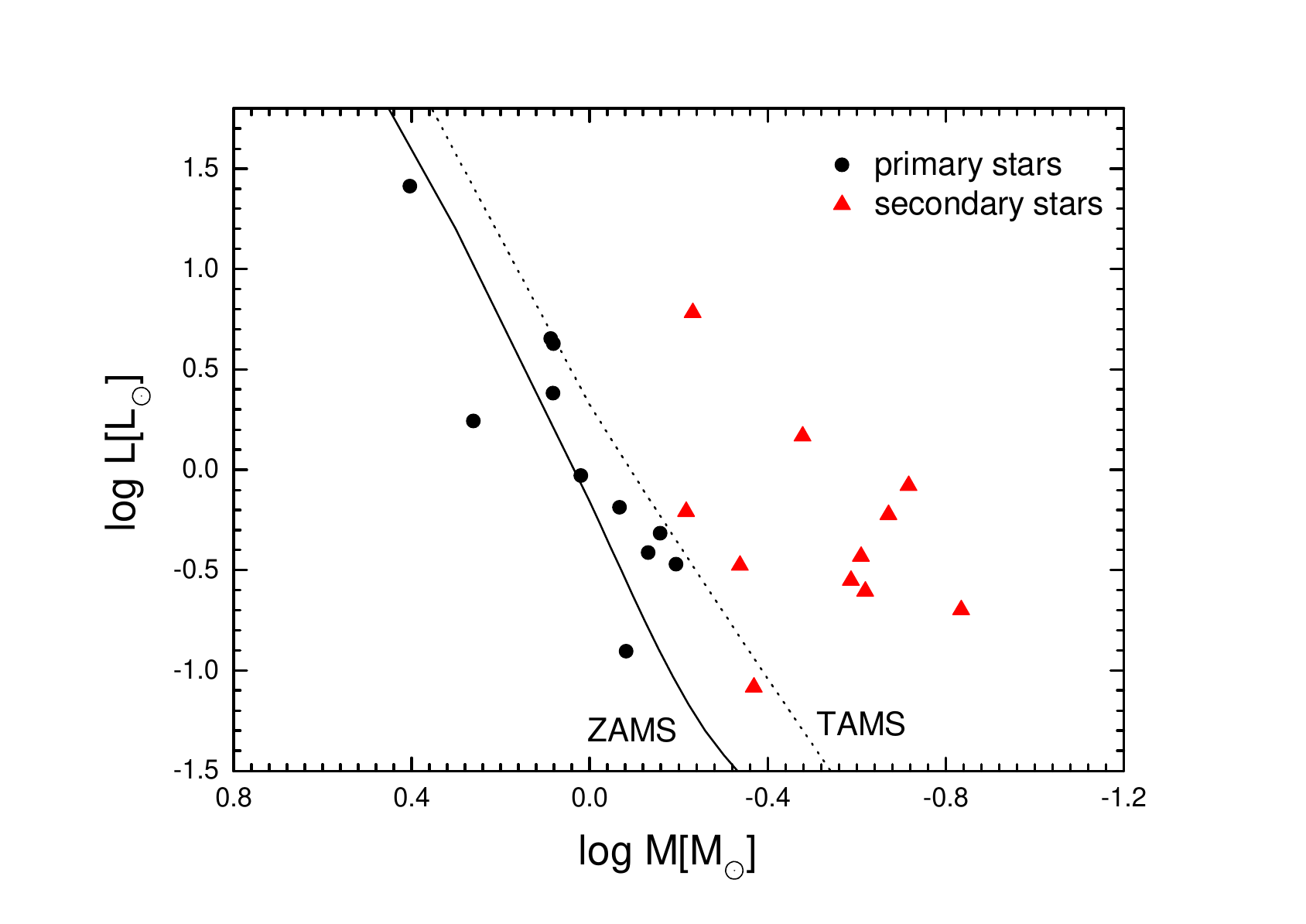}}
    \end{minipage}
    \hspace{-3.5cm} 
    \begin{minipage}{0.55\textwidth}
        \hspace{2cm} 
        \centerline{\includegraphics[width=\linewidth]{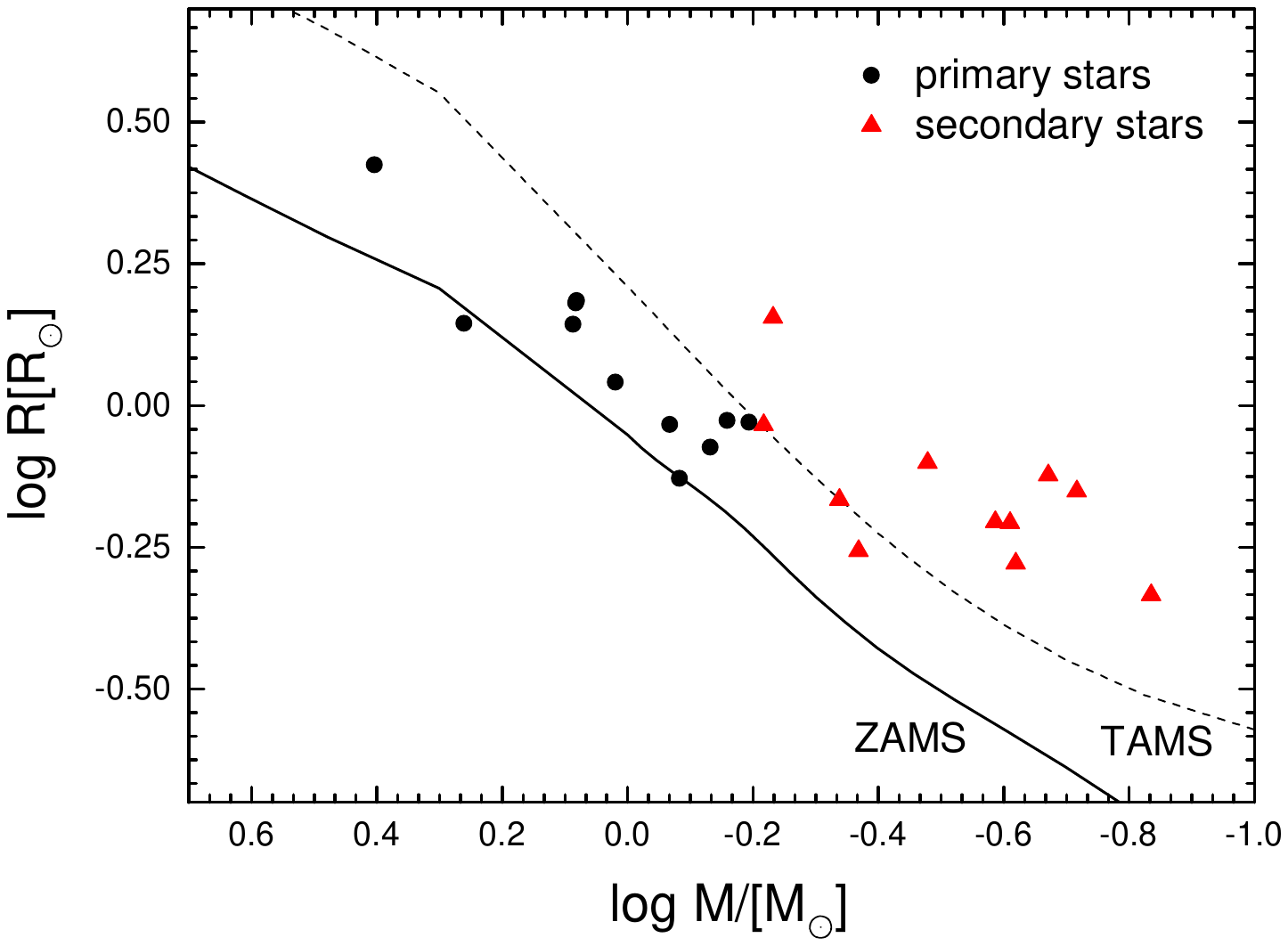}}
    \end{minipage}
    \caption{The M-L and M-R relations for 11 contact binaries.}
    \label{M-L-R}
\end{figure*}

In order to study the evolutionary states of these binary stars, the mass–radius (M-R) and the mass–luminosity (M-L) diagrams were plotted in Figure \ref{M-L-R}. ZAMS line refers to the zero age main sequence line, while TAMS line refers to the terminal age main sequence line, which are constructed by the binary evolutionary code provided by \cite{2002MNRAS.329..897H}. 
Solid dots represent the primary stars, and solid triangles represent the secondary stars. As can be seen from the figure, the primary star is located near the ZAMS, while most of the secondary stars are to the upper right of the TAMS. 
This indicates that, compared to the main sequence stars of the same mass, the secondary stars exhibit higher luminosities and radii. This phenomenon may be attributed to the transfer of mass and energy from the primary star to the secondary one\citep{1941ApJ....93..133K, 2022ApJS..262...12K}, or alternatively, it could result from the secondary stars having evolved out of the main sequence phase \citep{2013MNRAS.430.2029Y}.

\begin{figure*}
    \centering 
    \begin{minipage}{0.7\textwidth}
        \centerline{\includegraphics[width=\linewidth]{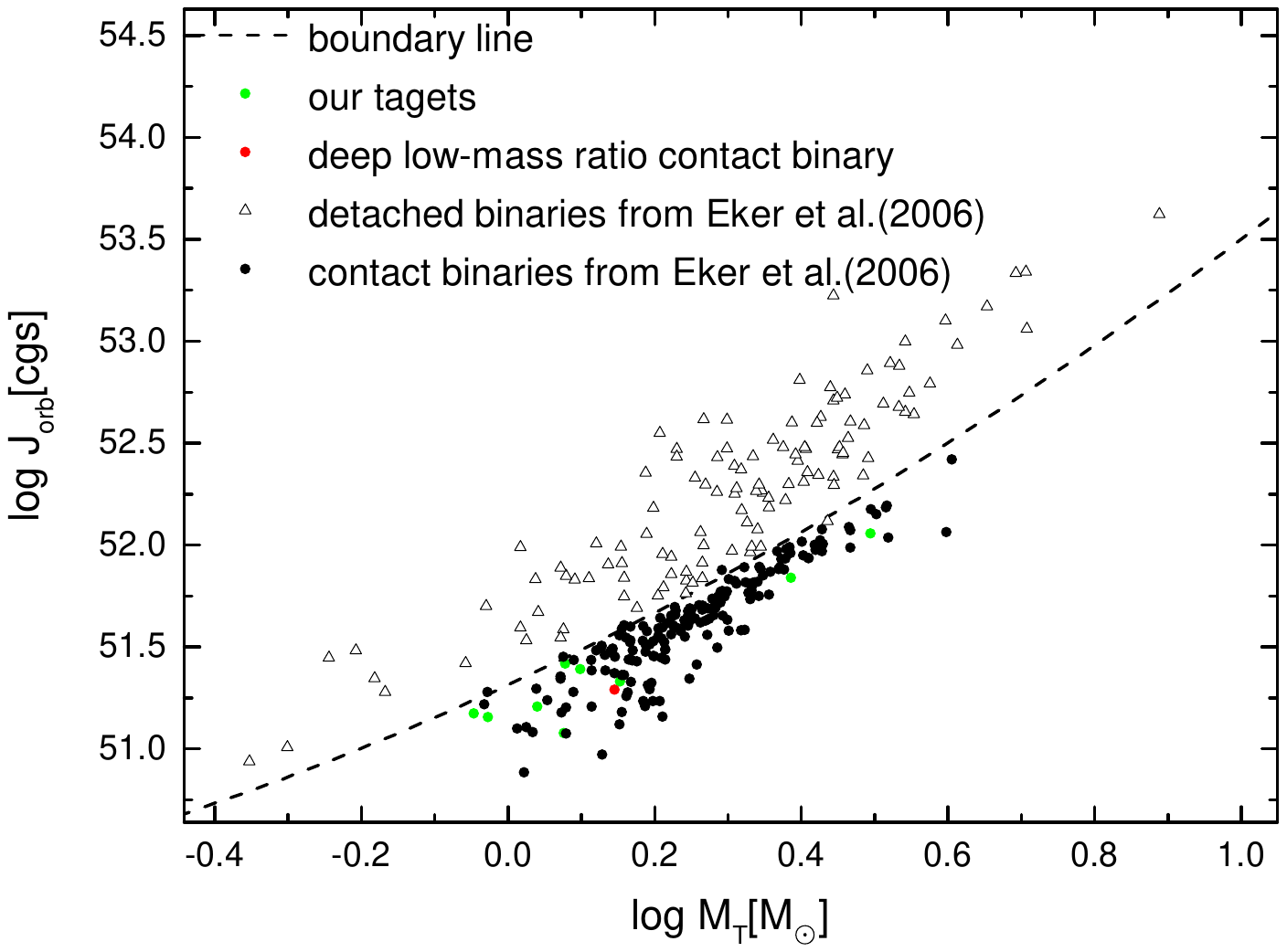}}
    \end{minipage}
    \caption{The relation between \( M_{T} \) and \( J_{\text{orb}} \).  Triangles and black dots represent detached binary stars and contact binary stars from \cite{2006MNRAS.373.1483E}, green dots denote our target contact binary stars, the red dots indicate deep low-mass ratio contact binary stars, and the dashed line represents the boundary between detached binary stars and contact binary stars from \cite{2006MNRAS.373.1483E}.}
    \label{M-J-11}
\end{figure*}

The orbital angular momentum was calculated according to the following equation provided by \cite{2006MNRAS.373.1483E}:

\begin{equation}
J_{orb}=1.24 \times 10^{52} \times M_{T}^{5/3} \times P^{1/3} \times q \times (1+q)^{-2},
\end{equation}
where $M_{T}$ refers to total mass of contact binary stars in solar units and the orbital period is in days, and the unit of the leading constant is erg·s.
These values are displayed in Table \ref{tab:solution}. The $J_{orb}$-$M_{T}$ diagram is displayed in Figure \ref{M-J-11}, with both the detached binary stars and the boundary between detached and contact binaries from \cite{2006MNRAS.373.1483E} plotted therein. As illustrated in the diagram, all our target objects lie below this boundary, residing within the contact binary region—consistent with the results in \cite{2006MNRAS.373.1483E}. The red dots represent low-mass ratio (q<0.25) deep (f>50\%) contact binary star TIC 142587827. Compared to contact binaries of the same mass, TIC 142587827 exhibits a lower orbital angular momentum, suggesting that it may be in the late evolutionary stage of contact binaries and may undergo merger in the subsequent evolutionary process. Therefore, TIC 142587827 is regarded as a candidate for contact binary merger.

In summary, we conducted spectral and photometric analyses on 11 targets, obtaining accurate physical parameters for these systems. Among them, there are 5 shallow-contact binaries, 3 moderate-contact binaries, and 3 deep-contact binaries. We identified a candidate for a binary merger. Additionally, we analyzed their evolutionary status and orbital changes.



\begin{acknowledgments}

This work was supported by National Natural Science Foundation of China (NSFC, Nos. 12273018 and 12573033), and by the Taishan Scholars Young Expert Program of Shandong Province, and by the Qilu Young Researcher Project of Shandong University, and by Shandong Provincial Natural Science Foundation of China (Nos. ZR2025MS81 and ZR2025MS90), and by Young Data Scientist Project of the National Astronomical Data Center and by the Cultiation Project for LAMOST Scientific Payoff and Research Achievement of CAMS-CAS, and by the Chinese Academy of Science Interdisciplinary Innovation Team. The calculations in this work were carried out at Supercomputing Center of Shandong University,Weihai. 

The data used in this paper comes from TESS mission. Funding for the TESS mission is provided by NASA Science Mission Directorate. We acknowledge the TESS team for its support of this work.

This work includes data collected by Guoshoujing Telescope (the Large Sky Area Multi-Object Fiber Spectroscopic Telescope LAMOST), which is a national major scientific project built by the Chinese Academy of Sciences. Funding for the project has been provided by the National Development and Reform Commission. 
LAMOST is operated and managed by the National Astronomical Observatories, Chinese Academy of Sciences. 

This work has made use of data from the European Space Agency (ESA) mission Gaia (https://www.cosmos.esa.int/gaia), processed by the Gaia Data Processing and Analysis Consortium (DPAC; https://www.cosmos.esa.int/web/gaia/dpac/consortium). Funding for the DPAC has been provided by national institutions, in particular the institutions participating in the Gaia Multilateral Agreement.
\end{acknowledgments}

\bibliography{sample631}
\bibliographystyle{aasjournal}

\end{document}